\documentclass[conference]{IEEEtran}

\usepackage{bm}
\usepackage{url}
\let\labelindent\relax
\usepackage{enumitem}
\usepackage[dvipsnames]{xcolor}
\usepackage{subcaption}
\usepackage{balance}  % for  \balance command ON LAST PAGE  (only there!)
\usepackage[linesnumbered,ruled,boxed,vlined]{algorithm2e}
\usepackage{graphicx}
\usepackage{epsfig}

\SetCommentSty{mycommfont}

\long\def\comment#1{}
\newcounter{example}[section]
\renewcommand{\theexample}{\nthesection.\arabic{example}}
\newenvironment{example}{
     \refstepcounter{example}
     {\vspace{1ex} \noindent\bf  Example  \theexample:}}{
     \vspace{1ex}} %\hspace*{\fill}\vspace*{1ex}}

\newcounter{definition}[section]
\renewcommand{\thedefinition}{\nthesection.\arabic{definition}}
\newenvironment{definition}{
     \refstepcounter{definition}
     {\vspace{1ex} \noindent\bf  Definition  \thedefinition:}}{
     \vspace{1ex}} %\hspace*{\fill}\vspace*{1ex}}

\newcounter{invariant}[section]
\renewcommand{\theinvariant}{\nthesection.\arabic{invariant}}
\newcounter{theorem}[section]
\renewcommand{\thetheorem}{\nthesection.\arabic{theorem}}
\newenvironment{theorem}{\begin{em}
        \refstepcounter{theorem}
        {\vspace{1ex} \noindent\bf  Theorem  \thetheorem:}}{
        \end{em}\vspace{1ex}} %\hspace*{\fill}\vspace*{1ex}}

\newcounter{lemma}[section]
\renewcommand{\thelemma}{\nthesection.\arabic{lemma}}
\newenvironment{lemma}{\begin{em}
        \refstepcounter{lemma}
        {\vspace{1ex}\noindent\bf Lemma \thelemma:}}{
        \end{em}\vspace{1ex}} %\hspace*{\fill}\vspace*{1ex}}

\newcounter{remark}[section]
\renewcommand{\theremark}{\nthesection.\arabic{remark}}
\newcommand{\proofsketch}{\noindent{\bf Proof: }}

\newcommand{\nthesection}{\arabic{section}}
\newcommand{\stitle}[1]{\vspace{1ex} \noindent{\bf #1}}

\newcommand{\kw}[1]{{\ensuremath {\mathsf{#1}}}\xspace}

\newcommand{\rem}{\kw{deg}^+}
\newcommand{\ext}{\kw{deg}^*}
\newcommand{\CD}{CoreDecomp\xspace}
\newcommand{\OI}{OrderInsert\xspace}
\newcommand{\OR}{OrderRemoval\xspace}
\newcommand{\TRA}{Trav\xspace}
\newcommand{\order}{\preceq}
\newcommand{\extdegree}{candidate degree\xspace}
\newcommand{\fastest}[1]{\color{red}{\textbf{#1}}}
\newcommand{\runnerup}[1]{\color{blue}{\textit{#1}}}

\newcommand{\NM}[1]{\textnormal{#1}}

\newcommand{\TT}[1]{\kw{#1}}

\newcounter{observation}[section]
\renewcommand{\theobservation}{\nthesection.\arabic{observation}}
\newenvironment{observation}{\begin{em}
        \refstepcounter{observation}
        {\vspace{1ex}\noindent\bf Observation \theobservation:}}{
  \end{em}\vspace{1ex}} %\hspace*{\fill}\vspace*{1ex}}

\begin{document}

% ****************** TITLE ****************************************

\title{A Fast Order-Based Approach for Core Maintenance}

\author{
{Yikai Zhang$^{\dag}$, Jeffrey Xu Yu$^{\dag}$, Ying Zhang$^{\ddag}$,
  and Lu Qin$^{\ddag}$}%  
\vspace{1.6mm}\\
\fontsize{10}{10}\selectfont\itshape
$^{\dag}$The Chinese University of Hong Kong, Hong Kong, China \\
\fontsize{10}{10}\selectfont\itshape
$^{\ddag}$Centre for QCIS, FEIT, University of Technology, Sydney, Australia\\
\fontsize{9}{9}\selectfont\ttfamily\upshape
\{ykzhang,yu\}@se.cuhk.edu.hk; \{Ying.Zhang,Lu.Qin\}@uts.edu.au
}

\maketitle

\begin{abstract}
  Graphs have been widely used in many applications such as social
  networks, collaboration networks, and biological networks. One
  important graph analytics is to explore cohesive subgraphs in a
  large graph.  Among several cohesive subgraphs studied, $k$-core is
  one that can be computed in linear time for a static graph. Since graphs
  are evolving in real applications, in this paper, we study core
  maintenance which is to reduce the computational cost to compute
  $k$-cores for a graph when graphs are updated from time to time
  dynamically. We identify drawbacks of the existing efficient
  algorithm, which needs a large search space to find the vertices
  that need to be updated, and has high overhead to maintain the index
  built, when a graph is updated. We propose a new order-based
  approach to maintain an order, called $k$-order, among vertices,
  while a graph is updated.  Our new algorithm can significantly
  outperform the state-of-the-art algorithm up to 3 orders of magnitude
  for the 11 large real graphs tested. We report our findings in this
  paper.
\end{abstract}

\section{Introduction} \label{sec:introduction}

Due to the ubiquity of graph data in different applications, graph
analytics has attracted much attention from both research and industry
communities. One major issue in graph analytics is to identify cohesive
subgraphs in the graphs, such as $k$-cores
\cite{batagelj:core-decomposition}, $k$-trusses \cite{wang:truss},
cliques, $n$-cliques, and $n$-clans \cite{cohen:trusses}.  Among them,
a $k$-core is defined as the maximal subgraph of an undirected graph
$G$ such that all vertices in the subgraph have degree of at least
$k$. For each vertex $v$ in $G$, the core number of $v$ is defined as
the maximum $k$ such that $v$ is contained in the $k$-core of
$G$. $k$-cores can be computed in linear time, whereas the time complexity
for $k$-trusses is $\mathcal{O}(m^{1.5})$ \cite{wang:truss}, and
cliques, $n$-cliques, and $n$-clans are NP-hard problems
\cite{cohen:trusses}.
%
% Given a graph $G$, In literature, the problem of computing core
% numbers for all vertices in $G$ is called \textit{core decomposition}
% and a linear in-memory algorithm \cite{batagelj:core-decomposition} is
% known to do the core decomposition efficiently.
%
Due to its linear computability, $k$-cores have been widely used in
many real-world applications, including graph visualization
\cite{alvarez:large}, community search \cite{li:efficient}, system
structure analysis \cite{zhang:using}, network topology analysis
\cite{alvarez:how} and so on.  There are many works studying how to
compute the core number for every vertex in a static graph
efficiently. Such a problem is known as a \textit{core decomposition}
problem~\cite{batagelj:core-decomposition,cheng:efficient,khaouid:kcore,montresor:distributed,wen:efficient}.

However, in many real-world applications, such as online social
network, collaboration network, and Internet, graphs are evolving
where vertices/edges will be inserted/removed over time dynamically.
There are increasing interests to study how to handle dynamic
graphs. To mention a few, \cite{zhu:reachability} proposes a general
technique to maintain a class of $2$-hop labels for reachability
queries in dynamic graphs; \cite{epasto:efficient} presents a scalable
algorithm to maintain approximate densest subgraphs in the dynamic
graph model with provable guarantee; \cite{fan:incremental} solves the
problem of incrementally maintaining the matches to a query pattern
when updates to the data graph are allowed.
The problem of maintaining core numbers for an evolving graph is
called \textit{core maintenance}. In brief, after inserting an edge into or
removing an edge from an undirected graph $G = (V, E)$, the problem is
how to update the core numbers for the vertices that need to be
updated.  There are two important tasks in a core maintenance
algorithm. First, it needs to identify the set of vertices whose core
numbers need to be updated when an edge is inserted/removed.
Such a set is denoted as $V^*$. Second, it recomputes the core
numbers for the vertices in $V^*$. In order to achieve high efficiency, an
index is used and the index needs to be maintained accordingly.  In
this paper, we study efficient in-memory core maintenance.  Ideally, a
good core maintenance algorithm should satisfy the following three criteria:
(a) a small performance variation among edge updates,
(b) a small cost for identifying
$V^*$, and (c) a small cost for updating the index.

\begin{figure}[t]
\centering
\epsfig{file=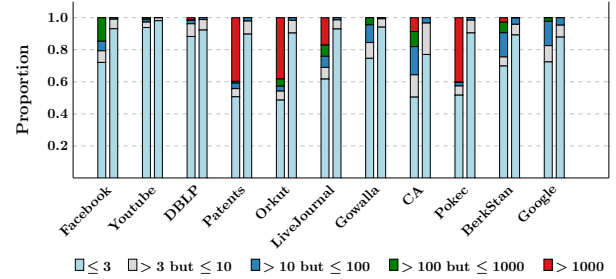,width=0.9\columnwidth}%,height=2.8cm}
\caption{The distribution of the number of vertices visited: the
traversal insertion algorithm (left bar) vs our new order-based
insertion algorithm (right bar)}
\label{fig:dist}
\end{figure}

\begin{figure}[t]
\centering
\epsfig{file=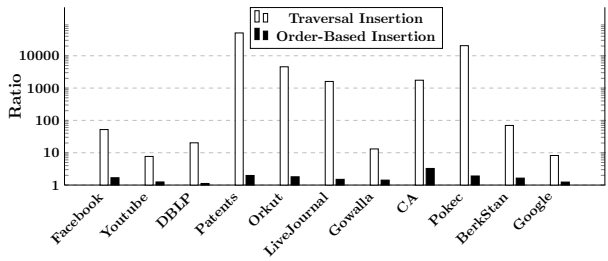,width=0.9\columnwidth}%,height=2.8cm}
\caption{Ratio of number of vertices visited: the traversal insertion
  algorithm vs our new order-based insertion algorithm}
\label{fig:perf-expansion}
\end{figure}

Sariy{\"u}ce et al. in \cite{sariyuce:streaming} propose an algorithm,
called traversal algorithm, which is the state-of-the-art approach for
core maintenance.  The traversal algorithm searches for $V^*$ only in
a local region near the edge inserted/removed. Therefore, it is much
faster than recomputing core numbers from scratch.  However, the
traversal algorithm has drawbacks.
First, it shows high variation in terms of performance when edges are
inserted into a graph.  We conduct testings to insert 100,000 edges
into 11 real graphs tested (Table~\ref{tbl:dataset}).  For each
edge inserted, the algorithm visits some vertices, denoted as $V'$, in order to
identify $V^*$.
We show the distribution of the size of $V'$ over all the 100,000 edges inserted
in \figurename~\ref{fig:dist}. As shown in \figurename~\ref{fig:dist}, for
each of the 11 real graphs tested, there are two bars. The left bar is by
the traversal insertion algorithm, whereas the right bar is by our new
order-based insertion algorithm.
A bar shows the corresponding proportion of edges inserted for different
$|V'|$.
For example, for a large proportion
of the 100,000 edges inserted, both our algorithm and the traversal algorithm
only visit up to 3 vertices. However, the traversal insertion algorithm needs
to visit more than 1,000 vertices for a non-small proportion of edge
insertions, as shown by certain left bars. Our order-based algorithm
only needs to visit up to 100 vertices for any of the 11 graphs, showing small
performance variation.
%%%%%%%%%%%%%%%%%%%%%%%%%%
%%%%%%%%%%%%%%%%%%%%%%%%%%%
Second, we show the ratio of the size of the set of vertices visited
($|V'|$) to the size of the set of vertices ($|V^*|$) that need to
be updated, i.e., {\small$\frac{\NM{sum of } |V'|\NM{ over all
    insertions}}{\NM{sum of }|V^*|\NM{ over all insertions}}$} for the
100,000 edge insertions over each of the same 11 graphs in
\figurename~\ref{fig:perf-expansion}.
For the
traversal insertion algorithm, the ratio is at least $7$ and can be
up to 10,000 for \TT{Patents} and \TT{Pokec}. It suggests the deficiency of the
traversal insertion algorithm, even though it is the state-of-the-art.
In contrast, for our new insertion algorithm, the ratios are below $4$
for all graphs tested and can be down to about 1.
Third, for the traversal algorithm, the cost to maintain the index is high. The
traversal algorithm maintains a value $\TT{pcd}(v)$, called the pure-core degree of $v$,
for each $v$. Here, $\TT{pcd}(v)$ is computed from the core
numbers of all vertices in the $2$-hops neighborhood of $v$.
Therefore, if the core number of a vertex changes, a large
number of vertices may have their $\TT{pcd}$ values to be updated,
which incurs high cost. We will revisit the traversal algorithm in
details in Section \ref{sec:traversal}.

The main contributions of this work are summarized below. First, we
identify the deficiency of the state-of-the-art traversal algorithm
\cite{sariyuce:streaming} for core maintenance. Second, we propose new
order-based in-memory algorithms for core maintenance:
\TT{\OI} for edge insertion and \TT{\OR} for edge removal. Note that,
in this work, we do not consider vertex insertion/removal, since they
can be simulated by a sequence of edge insertions and removals
\cite{li:efficient,sariyuce:streaming,wen:efficient}.
The main idea behind our new order-based algorithms is that we
explicitly maintain a $k$-order among vertices such that $u \order v$
for every two vertices in a graph $G$.  Here, a $k$-order, $(v_1, v_2,
\ldots, v_n)$, for every vertex $v_i$ in a graph $G$, is an instance
of all the possible vertex sequences produced by a core decomposition
algorithm~\cite{batagelj:core-decomposition}. In other words, there
are many possible vertex sequences produced by a core decomposition
algorithm, and we ensure the $k$-order under maintenance is one of them at any
time for any evolving graph $G$.  The transitivity holds. That is, $v_h \order
v_j$ if $v_h \order v_i$ and $v_i \order v_j$.  Given a $k$-order
maintained for $G$, when an edge is inserted/removed, we maintain the
$k$-order for the new $G$ with the edge inserted/removed.
%%%%%%%%%%%%%%%%%%%%%%%%%%%%%%%%%%%%%%%%%%%%%%%
Third, we have conducted extensive performance studies, and show that
we can achieve high efficiency up to 2,083 times faster than the
traversal insertion algorithm.

The rest of this paper is organized as follows. We discuss the related
work in Section \ref{sec:related}. We give the preliminaries in
Section \ref{sec:background}, and review the traversal
insertion/removal algorithm in Section \ref{sec:traversal}. We discuss
our new order-based algorithm, namely, \TT{\OI} and \TT{\OR}, in
Section \ref{sec:basic-algs}, and give some implementation details in
Section \ref{sec:list}. We conduct extensive performance studies, and
report the results in Section \ref{sec:exp}, and conclude this work in
Section \ref{sec:con}.

\section{Related Work} \label{sec:related}

Identifying cohesive subgraphs is an important problem in graph analytics.
Common cohesive subgraphs are clique (maximal clique), $k$-plex, $n$-clique,
$n$-clan \cite{cohen:trusses}, $k$-truss \cite{wang:truss}, $k$-core
\cite{batagelj:core-decomposition} and so on, among which $k$-core is the
only one known to have a linear algorithm. We review the related work
on computing $k$-core (core decomposition) and maintaining $k$-core (core
maintenance) respectively below.

\stitle{Core Decomposition:} The core decomposition is to efficiently
compute for each vertex its core number.
\cite{batagelj:core-decomposition} proposes an in-memory linear
algorithm ($\mathcal{O}(m+n)$).  Their algorithm uses a bottom-up
process in which $k$-cores are computed in the order of $1$-core,
$2$-core, $3$-core, $\cdots$.  To process graphs that can not reside
in the memory, \cite{cheng:efficient} proposes an external algorithm,
which runs in a top-down manner such that the whole graph does not
need to be loaded to memory to compute higher
cores. \cite{wen:efficient} proposes a new semi-external algorithm,
which assumes that the memory has at least $\Omega(n)$ size and
can maintain a small constant amount of information for each vertex in
memory.  \cite{montresor:distributed} considers core decomposition
under a distributed setting.  \cite{khaouid:kcore} investigates core
decomposition of large graphs using \TT{GraphChi}, \TT{WebGraph}, and external
model, and compares their performance in a single PC.

\stitle{Core Maintenance:} Given a large graph, it takes high
computational cost to recompute core numbers for vertices when some
edges are inserted/removed, even though there exist linear algorithms.
The core maintenance is to maintain core numbers efficiently when edges
and vertices are inserted/removed. Since the insertions and removals
of vertices can be simulated by a sequence of edge insertions and
removals, all existing works
\cite{li:efficient,sariyuce:streaming,wen:efficient,aksu:distributed} consider
only edge updates. The key issue is how to efficiently identify the set of
vertices, denoted as $V^*$, whose core numbers need to be updated, around the
edge inserted/removed.
% 
% 
% Rather than doing core decomposition again from scratch,
% it is more expected that a core maintenance algorithm can work
% \textit{locally}, i.e., only visit and expand vertices that are near
% $V^*$, which is the set of vertices whose core numbers need to be
% updated.
%
Recently, \cite{li:efficient} and \cite{sariyuce:streaming}
independently found that, when an edge is inserted or removed, the
induced subgraph of $V^*$ is connected and moreover, $V^*$
resides in the subcore which the edge is in.  Based on this
observation, \cite{sariyuce:streaming} proposes algorithms that are linear in
the size of the subcore, whereas \cite{li:efficient} proposes a
quadratic algorithm. Among the algorithms proposed in
\cite{sariyuce:streaming}, the traversal algorithm is the
state-of-the-art for core maintenance.
%
% For one thing, it precomputes
% $\TT{mcd}$ and $\TT{pcd}$ values to reduce search space and reduce
% computational cost in runtime; for another, it uses a ``root aware''
% mechanism to early terminate the algorithm.
%
We will revisit the traversal algorithm later, and show that we can
significantly improve the efficiency when a graph can be held in
memory.  A semi-external algorithm for core maintenance is
proposed in \cite{wen:efficient}
to reduce I/O cost; but it is not optimized for CPU time. The algorithm proposed
by \cite{aksu:distributed} is similar to the \TT{SubCore} algorithm of \cite{sariyuce:streaming} but is
less efficient due to weaker bounds.

\section{The Preliminaries} \label{sec:background}

Let $G = (V, E)$ be an undirected graph, where $V(G)$ denotes the set
of vertices and $E(G)$ represents the set of edges in $G$. We denote
the number of vertices and edges of $G$ by $n$ and $m$
respectively. In this paper, we use $\TT{nbr}(u, G)$ to denote the set
of neighbors of a vertex $u \in V(G)$, i.e., $\TT{nbr}(u, G) = \{v \in
V(G): (u, v) \in E(G)\}$. Besides, we define the degree of $u$ in $G$
as $\TT{deg}(u, G) = |\TT{nbr}(u, G)|$. When the context is clear, we
will use $\TT{nbr}(u)$ and $\TT{deg}(u)$ instead of $\TT{nbr}(u, G)$
and $\TT{deg}(u, G)$.  We say a graph $G'$ is a subgraph of $G$,
denoted as $G' \subseteq G$, if $V(G') \subseteq V(G)$ and $E(G')
\subseteq E(G)$. Given a subset $V' \subseteq V$, the subgraph induced
by $V'$, denoted as $G(V')$, is defined as $G(V') = (V', E')$ where
$E' = \{(u, v) \in E: u, v \in V'\}$.

%\begin{definition}\label{def:kcore}
A subgraph $G_k$ of $G$ is called a \textbf{$\bm{k}$-core} if it satisfies
the following conditions: (1) for $\forall u \in V(G_k)$, 
$\TT{deg}(u, G_k) \geq k$; (2) $G_k$ is maximal.
$G_k = \emptyset$ if the $k$-core of $G$ does not exist.
%\end{definition}
%
For a given $k$, the $k$-core $G_k$ of a graph $G$ is unique.
Moreover,
% \begin{equation} \label{eq:core-inclusion}
$G_{k+1} \subseteq G_k, \textnormal{ for } \forall k \geq 0$.
% \end{equation}
Note that when $k = 0$, $G_0$ is just $G$. A closely related concept to $k$-core
is \textbf{core number}, which is defined as follows:
%
%\begin{definition}\label{def:core-number}
For each vertex $u \in V(G)$, its core number $\TT{core}(u,$ $G)$ is defined
as $\TT{core}(u, G) = \max\{k: u \in V(G_k)\}$.
When the context is clear, for simplicity, we
use $\TT{core}(u)$ to denote the core number of $u$ instead.
%\end{definition}
%
%\begin{definition}\label{def:subcore}
Given a graph $G = (V, E)$, a set of vertices $\mathcal{SC} \subseteq V$ is
called a \textbf{$\bm{k}$-subcore} if (1) $\forall u \in \mathcal{SC}$,
$\TT{core}(u) = k$; (2) the induced subgraph $G(\mathcal{SC})$ is connected; (3)
$\mathcal{SC}$ is maximal. For a vertex $u$, the subcore containing $u$ is
denoted as $\TT{sc}(u)$.
%\end{definition}

\begin{figure}[t]
\centering
\epsfig{file=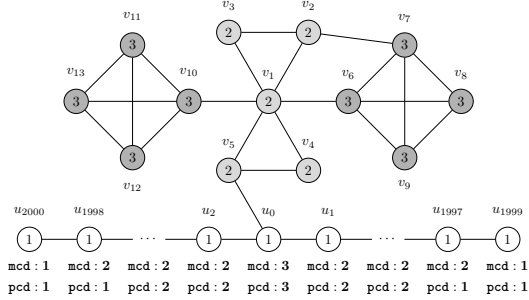,scale=0.53}
\vspace{-0.2cm}
\caption{A Sample Graph $G$.} \label{fig:example}
\end{figure}

\begin{example} \label{exm:core}
Consider the graph $G$ in \figurename~\ref{fig:example}. The subgraph
induced by $\{v_6, v_7, \cdots, v_{13}\}$ is the $3$-core of $G$
because in the induced graph, every vertex has a degree at least $3$.
There does not exist a $4$-core in $G$. We have $\TT{core}(v_i) = 3$ for
$6 \leq i \leq 13$, $\TT{core}(v_i) = 2$ for $1 \leq i \leq 5$ and
$\TT{core}(u_i) = 1$ for $0 \leq i \leq 2000$. Also, $\{v_1, v_2, v_3,
v_4, v_5\}$ and $\{u_i \}$ are the unique $2$-subcore and $1$-subcore
of $G$, respectively. There are two $3$-subcores in $G$ by $\{v_6,
v_7, v_8, v_9\}$ and $\{v_{10}, v_{11}, v_{12}, v_{13}\}$.
\end{example}

\stitle{Core Decomposition}: Given a graph $G$, the problem of
computing the core number for each $u \in V(G)$ is called core
decomposition.
\cite{batagelj:core-decomposition} proposed an $\mathcal{O}(m + n)$
algorithm, denoted as \TT{\CD}, whose sketch is shown in Algorithm
\ref{alg:core-decomposition}.  The general idea is as follows: To
compute the $k$-core $G_k$ of $G$, it repeatedly remove those vertices
(and their adjacent edges) whose degrees are less than $k$. When there
is no more vertex to remove, the resulting graph is the $k$-core of
$G$.

\stitle{Core Maintenance}: The problem of core maintenance is to
maintain the $k$-core by maintaining the core numbers, when edges are
inserted to or removed from $G$. This is because it is known that
$k$-core $G_k$ can be efficiently computed using the core numbers of
the vertices, since a vertex $u \in V(G_k)$ if and only if
$\TT{core}(u) \geq k$. In addition, an index can be constructed
for core numbers of the vertices.
From now on, we use the notation $V^*$ to denote the set of vertices
whose core numbers need to be updated after inserting or removing an
edge. The insertion or removal of vertices can be simulated as a sequence
of edge insertions and removals. Hence in this paper, we focus on efficient
core maintenance under edge updates.

\begin{algorithm}[t]
{\small
\linespread{0.8}\selectfont
\SetKwInOut{Input}{input} \SetKwInOut{Output}{output}
\SetKw{SuchThat}{such that}
$k \leftarrow 1$\;
\While{$G$ is not empty}{
	\While{$\exists u \in V(G)$ \SuchThat $\TT{deg}(u) < k$}{
		let $u$ be a vertex with $\TT{deg}(u) < k$\;
		$\TT{deg}(w) \leftarrow \TT{deg}(w) - 1$ for each $w \in \TT{nbr}(u)$\;
		remove $u$ and its adjacent edges from $G$\;
		$\TT{core}(u) \leftarrow k - 1$\;
		\label{line:k-order-gen}
	}
	$k \leftarrow k + 1$\;
}
\Return \ \TT{core}\;
}
\caption{\TT{\CD}($G$)} \label{alg:core-decomposition}
\end{algorithm}

Here we present two theorems given in
\cite{li:efficient,sariyuce:streaming} based on which the correctness of our
algorithms is proved.

\begin{theorem}\NM{\ \cite{li:efficient,sariyuce:streaming}}
\label{thm:core-diff}
After inserting to (resp. removing from) $G = (V, E)$ an edge, the core number
of a vertex $u \in V$ increases (resp. decreases) by at most $1$.
\end{theorem}

\begin{theorem}\NM{\ \cite{li:efficient,sariyuce:streaming}}
\label{thm:core-region}
Suppose an edge $(u, v)$ with $K = \TT{core}(u) \leq \TT{core}(v)$ is
inserted to (resp. removed from) $G$. Suppose $V^*$ is non-empty. We
have the following:  (1) if $\TT{core}(u) < \TT{core}(v)$, then $u \in
V^*$ and $V^* \subseteq \TT{sc}(u)$; (2) if $\TT{core}(u) =
\TT{core}(v)$, then both $u$ and $v$ are in $V^*$ (resp. at least one
of $u$ and $v$ is in $V^*$) and $V^* \subseteq \TT{sc}(u) \cup
\TT{sc}(v)$; (3) the induced subgraph of $V^*$ in $G\cup\{(u, v)\}$
(resp. $G$) is connected.
\end{theorem}

\vspace{-0.4cm}
Theorem \ref{thm:core-diff} suggests that we only need to find $V^*$,
the set of vertices whose core numbers need to be updated. Once $V^*$
is found, we can then increase (or decrease) $\TT{core}$ for vertices
in $V^*$ by $1$ accordingly.
Theorem \ref{thm:core-region} suggests two things. (1) Only vertices
$w$ with $\TT{core}(w) = K$ \textit{may} be in $V^*$, i.e.,
$\TT{core}(w)$ may need to be updated. (2) We can search for $V^*$ in
a small local region near the inserted (or removed) edge (i.e., the
subcores containing $u$ and $v$), rather than in the whole $G$.

\section{The Traversal Algorithm} \label{sec:traversal}

In this section, we discuss the state-of-the-art core maintenance
solution, called the traversal algorithm \cite{sariyuce:streaming},
which is designed on top of two important notions, namely
$\TT{mcd}(u)$ and $\TT{pcd}(u)$. We introduce them below.

The \textbf{max-core degree} of a vertex $u$, denoted as $\TT{mcd}(u)$, is
defined as the number of $u$'s neighbors, $w$, such that $\TT{core}(w) \geq
\TT{core}(u)$.
Intuitively, $\TT{mcd}(u)$ counts $u$'s neighbors in the $(\TT{core}(u))$-core.
By the definition of $k$-core, $\TT{mcd}(u) \geq \TT{core}(u)$.
The \textbf{pure-core degree} of a vertex $u$, denoted as $\TT{pcd}(u)$, is
defined as the number of $u$'s neighbors, $w$, such that either
$\TT{core}(w) = \TT{core}(u)$ and $\TT{mcd}(w) >
\TT{core}(w)$ or $\TT{core}(w) > \TT{core}(u)$.
The main difference between $\TT{pcd}$ and
$\TT{mcd}$ is that $\TT{pcd}(u)$ further excludes neighbors $w$ with
$\TT{core}(w) = \TT{core}(u)$ and $\TT{mcd}(w) = \TT{core}(w)$. Thus
$\TT{pcd}(u) \leq \TT{mcd}(u)$.  We show $\TT{mcd}(u_i)$ and
$\TT{pcd}(u_i)$, for $\{u_i\}$, in \figurename~\ref{fig:example}
and an example below which demonstrates why $\TT{mcd}$ and $\TT{pcd}$ are
useful.

\begin{example} \label{exm:mcd}
%
% As an example, $\TT{mcd}(u_i)$ for $0 \leq i \leq 2000$ are shown in
% \figurename~\ref{fig:example}. 
%
Suppose edge $(v_4, u_0)$ is inserted to $G$ in
\figurename~\ref{fig:example}.  Due to the newly inserted edge,
both $\TT{mcd}(u_0)$ and $\TT{pcd}(u_0)$ become $4$. 
By Theorem \ref{thm:core-region}, only
core numbers of vertices in $\{u_i\}$ may need to be updated.
We show two cases. First, 
consider $u_{1999}$. $u_{1999}$ can not be in the
$2$-core of $G \cup \{(v_4, u_0)\}$, because the number of its neighbors in the
new $2$-core is upper bounded by $\TT{mcd}(u_{1999}) < 2$.
To see why $\TT{mcd}(u_{1999})$ is an upper bound, recall that
by Theorem~\ref{thm:core-diff} and Theorem~\ref{thm:core-region},
only vertices with core numbers greater than or equal to $1$ may be in the new
$2$-core.
%observe that
%$u_{1997}$ is the only neighbor of $u_{1999}$ that satisfies
%$\TT{core}(u_{1997}) \geq \TT{core}(u_{1999}) = 1$.
%
Second, consider $u_{1997}$. Regarding $\TT{mcd}(u_{1997})$, the max
possible number of $u_{1997}$'s neighbors in the new $2$-core is
$\TT{mcd}(u_{1997}) = 2$, so $\TT{core}(u_{1997})$ \textit{may}
need to be updated.
%In other words, $u_{1997}$ may be in the $2$-core
%of $G \cup \{(v_4,u_0)\}$, because $\TT{core}(u_{1997})$ may become
%$2$.  This is supported by Theorem~\ref{thm:core-diff} and
%Theorem~\ref{thm:core-region}. 
Regarding $\TT{pcd}(u_{1997})$, the
max possible number of $u_{1997}$'s neighbors in the new $2$-core is
$\TT{pcd}(u_{1997})=1<2$. 
$\TT{pcd}(u_{1997})$ does not count $u_{1999}$, which can not be in
the new $2$-core because $\TT{mcd}(u_{1999}) = \TT{core}(u_{1999}) = 1$.
Hence, with $\TT{pcd}(u_{1997})$, there is
no need to update $\TT{core}(u_{1997})$.
\end{example}

\subsection{The Traversal Insertion Algorithm}

The insertion algorithm employs an expand-shrink framework to determine
$V^*$, the set of vertices whose core numbers need to be updated. It works as
follows:
When a new edge $(u, v)$ with $K = \TT{core}(u) \leq \TT{core}(v)$ is
inserted into $G$, it first updates $\TT{mcd}$ and $\TT{pcd}$
accordingly based on the old $\TT{core}$ values.
Because $\TT{core}(u) \leq \TT{core}(v)$, it selects $u$ as the
root. Based on Theorem \ref{thm:core-region} and the discussion above,
in order to find $V^*$, it issues a DFS starting from the root and only visits
$w$ with $\TT{core}(w) = K$ and
$\TT{mcd}(w) > K$.
The algorithm maintains a value $\TT{cd}(w)$ for every vertex $w$
visited, where $\TT{cd}(w)$ represents the max possible number of
$w$'s neighbors in the new $(K + 1)$-core. Initially $\TT{cd}(w) =
\TT{pcd}(w)$.
During the DFS, it stops the search when the vertex $w$ visited is
confirmed to be $w \notin V^*$ (i.e., $\TT{cd}(w) \leq K$).  It evicts
$w$ (i.e., mark $w \notin V^*$) and decreases the $\TT{cd}$ values of
its neighbors by $1$.  In addition, it conducts a backward search from
$w$ to find more visited vertices to evict. Such a process is called
an eviction propagation. After the eviction propagation ends, it will
continue DFS in the original order, skipping $w$. When the algorithm
terminates, it is guaranteed that all visited but not evicted vertices
constitute $V^*$. As the final step, it needs to update $\TT{mcd}$ and
$\TT{pcd}$ for later insertions and removals.

\begin{figure}[t]
\centering
\epsfig{file=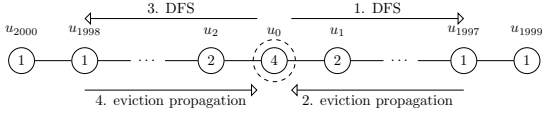, scale=0.56}
\vspace{-0.2cm}
\caption{Illustration of Example \ref{exm:traversal-ins}}
\label{fig:traversal-ins}
\end{figure}

\begin{example} \label{exm:traversal-ins}
Suppose $(v_4, u_0)$ is inserted to the graph $G$ in
\figurename~\ref{fig:example}. The initial $\TT{cd}$ values for all $u_i$ are
shown in \figurename~\ref{fig:traversal-ins}. The traversal insertion
algorithm selects $u_0$ as the root and issues a DFS from $u_0$. The
DFS recursively visits those vertices $u_i$ with $\TT{core}(u_i) = 1$ and
$\TT{mcd}(u_i) > 1$. Without loss of generality, assume the DFS first
traverses the right part of $u_0$. When DFS
reaches $u_{1997}$, it finds $\TT{cd}(u_{1997}) \leq 1$, thus
$\TT{core}(u_{1997})$ does not need to be updated and $u_{1997}$ is
evicted.  Further, the eviction propagates back to $u_1$ as follows:
Due to eviction of $u_{1997}$, $\TT{cd}(u_{1995})$ is decreased by $1$
and becomes $1$. Thus $u_{1995}$ is also evicted. This process continues
and finally it evicts all $u_{1997}$, $\cdots$,
$u_1$. The DFS then traverses the left part.  Similarly, $u_2$,
$\cdots$, $u_{1998}$ are all evicted. At last, $u_0$ is the only
vertex that is visited but not evicted and therefore $V^* = \{u_0\}$.
\end{example}

In Example~\ref{exm:traversal-ins}, although $u_0$ is the only vertex
whose core number needs to be updated, the traversal insertion
algorithm needs to visit much more (specifically,
1,999) vertices to determine $V^* = \{u_0\}$.
As shown in \figurename~\ref{fig:dist} and
\figurename~\ref{fig:perf-expansion}, such an issue is not rare and
hurts the performance significantly.

Such a high overhead is due to the potential large search space used
by the algorithm.  Let $V'$ be the search space, i.e., the set of
vertices that the traversal insertion algorithm visits, in
order to obtain $V^*$. In order to show an upper bound of $|V'|$, we
introduce a concept called \textit{pure-core}
\cite{sariyuce:streaming}, in Definition~\ref{def:purecore}.

\begin{definition} \label{def:purecore}
For each $u \in V(G)$, its \textbf{pure-core}, denoted as $\TT{pc}(u)$, is
defined as $\TT{pc}(u) = \{u\} \cup \mathcal{PC}$, where
$\mathcal{PC}$ is a set of vertices that satisfies the following 3
conditions: (1) for $\forall w \in \mathcal{PC}$, $\TT{mcd}(w) >
\TT{core}(w)$ and $\TT{core}(w) = \TT{core}(u)$; (2) $G(\{u\} \cup
\mathcal{PC})$ is connected; and (3) $\mathcal{PC}$ is maximal.
\end{definition}

\comment{
\begin{lemma} \label{lmm:purecore}
  Suppose $(u, v)$ is the newly inserted edge. (1) If $\TT{core}(u) <
  \TT{core}(v)$, then $V' \subseteq \TT{pc}(u) \subseteq \TT{sc}(u)$.
  (2) If $\TT{core}(u) = \TT{core}(v)$, then $V' \subseteq \TT{pc}(u)
  \cup \TT{pc}(v) \subseteq \TT{sc}(u) \cup \TT{sc}(v)$.
\end{lemma}
}

\begin{figure}[t]
% \begin{subfigure}{0.45\columnwidth}
% \epsfig{file=fig/facebook_search_space-figure0.eps,scale=0.4}
% \caption{Facebook}
% \end{subfigure}
% \begin{subfigure}{0.45\columnwidth}
% \epsfig{file=fig/youtube_search_space-figure0.eps,scale=0.4}
% \caption{Youtube}
% \end{subfigure}
% \\
% \begin{subfigure}{0.45\columnwidth}
% \epsfig{file=fig/dblp_search_space-figure0.eps,scale=0.4}
% \caption{DBLP}
% \end{subfigure}
\begin{subfigure}{0.49\columnwidth}
\centering
\epsfig{file=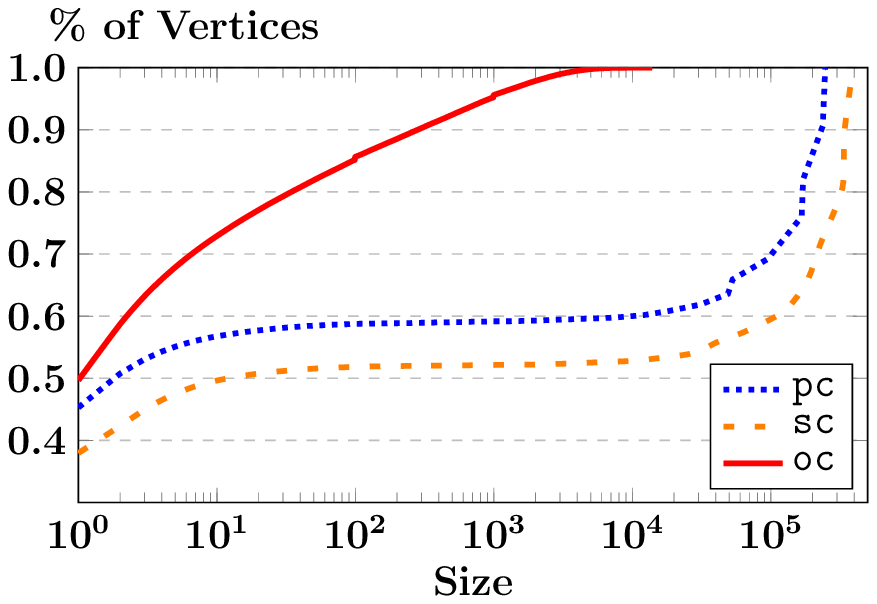,width=0.9\columnwidth}
\caption{\TT{Patents}}
\end{subfigure}
\begin{subfigure}{0.49\columnwidth}
\centering
\epsfig{file=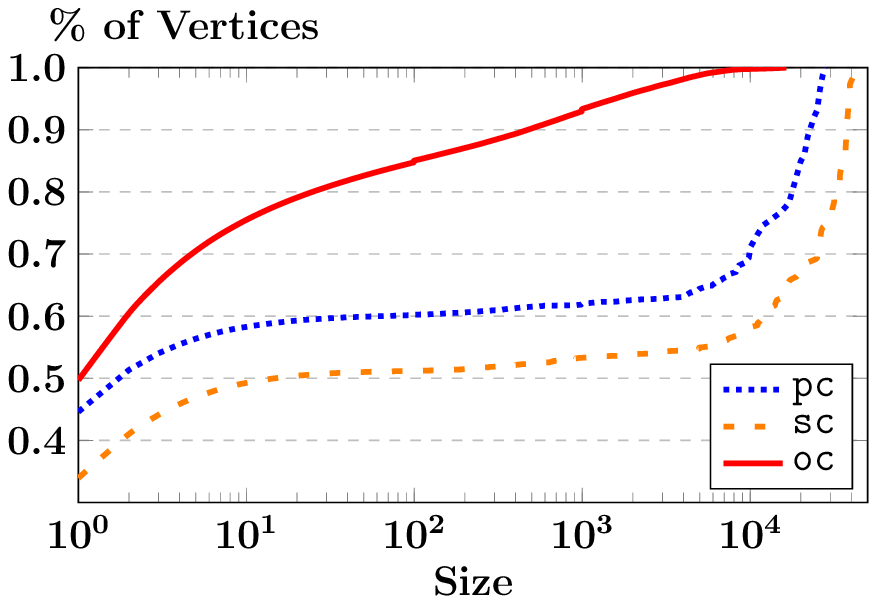,width=0.9\columnwidth}
\caption{\TT{Orkut}}
\end{subfigure}
% \begin{subfigure}{0.45\columnwidth}
% \epsfig{file=fig/livejournal_search_space-figure0.eps,scale=0.4}
% \caption{LiveJournal}
% \end{subfigure}
% \\
% \begin{subfigure}{0.45\columnwidth}
% \epsfig{file=fig/gowalla_search_space-figure0.eps,scale=0.4}
% \caption{Gowalla}
% \end{subfigure}
% \begin{subfigure}{0.45\columnwidth}
% \epsfig{file=fig/CA_search_space-figure0.eps,scale=0.4}
% \caption{CA}
% \end{subfigure}
% \\
% \begin{subfigure}{0.45\columnwidth}
% \epsfig{file=fig/pokec_search_space-figure0.eps,scale=0.4}
% \caption{Pokec}
% \end{subfigure}
% \begin{subfigure}{0.45\columnwidth}
% \epsfig{file=fig/berkstan_search_space-figure0.eps,scale=0.4}
% \caption{BerkStan}
% \end{subfigure}
% \\
% \begin{subfigure}{0.45\columnwidth}
% \epsfig{file=fig/google_search_space-figure0.eps,scale=0.4}
% \caption{Google}
% \end{subfigure}
\vspace{-0.2cm}
\caption{Cumulative size distribution of $\TT{pc}$, $\TT{sc}$, and $\TT{oc}$.}
\label{fig:size_dist}
\end{figure}

It is not hard to observe that $V' \subseteq \TT{pc}(u) \cup \TT{pc}(v)$
when an edge $(u, v)$ is inserted; that is, the
size of pure-cores decides the max number of vertices that are visited by the traversal insertion
algorithm. We show the cumulative distribution of $\TT{pc}$, for
two of the largest graphs tested (Table~\ref{tbl:dataset}), namely, \TT{Patents}
and \TT{Orkut}, in \figurename~\ref{fig:size_dist}.
%
% where the x-axis is the size
% $s$ (the number of vertices to be visited) and the y-axis is the \% of
% the such vertices).
%
\figurename~\ref{fig:size_dist} shows the \% of vertices whose pure-cores
($\TT{pc}$) are smaller than a certain size $s$ for all possible $s$
in the x-axis. We observe that $\TT{pc}$ is with high variance in
distribution. Moreover, many vertices have a very large pure-core
size. For example, for both graphs, more than 10\% of vertices have
their pure-core size greater than 10,000, which is very high and leads
to high overhead.
%
% As we will show later, $k$-orders
% help reduce the search space a lot and thus bring large improvement in
% performance.
%

\subsection{The Traversal Removal Algorithm} \label{sec:traversal-remove}

The removal algorithm first updates $\TT{mcd}$ and $\TT{pcd}$ when
an edge $(u, v)$ with $K = \TT{core}(u) \leq \TT{core}(v)$ is removed from $G$.
The algorithm uses the similar idea
as shown in the \TT{\CD} algorithm (Algorithm
\ref{alg:core-decomposition}) to find $V^*$.
%
% This is the core part of the traversal removal algorithm.
%
It works as follows. Let $u$ be the root (if $\TT{core}(v) = K$, $v$
should also be a root). Starting from the root(s), the algorithm finds $V^*$ by
repeatedly removing vertices $w$ with $\TT{core}(w) = K$ and $\TT{cd}(w) < K$ and
appending $w$ to $V^*$, where $\TT{cd}(w)$ is initialized as
$\TT{mcd}(w)$ and counts the max possible \# of neighbors of
$w$ in the new $K$-core.
When $w$ is removed,
the algorithm decreases $\TT{core}(w)$ by 1.  Since
$\TT{core}(w)$ becomes $K - 1$, for $z \in \TT{nbr}(w)$ with $\TT{core}(z) = K$,
$\TT{cd}(z)$ is decreased by $1$. Accordingly, if $\TT{cd}(z)$ becomes smaller
than $K$, $z$ is also removed.
Such a process continues until no more vertex can be removed and requires only
$\mathcal{O}(\sum_{v \in V^*}\TT{deg}(v))$ time. At last, $\TT{mcd}$ and $\TT{pcd}$ need to
be updated for later insertions and removals.

\comment{
\begin{example}
Suppose $(v_{10}, v_{11})$ is removed from $G$. Consequently, both
$\TT{mcd}(v_{10})$ and $\TT{mcd}(v_{11})$ decrease to $2$. By Theorem
\ref{thm:core-region}, only vertices in $\{v_{10}, v_{11}, v_{12},
v_{13}\}$ may have their core numbers to be updated. Their $\TT{cd}$
values are initially $\TT{cd}(v_{10}) = \TT{cd}(v_{11}) = 2$ and
$\TT{cd}(v_{12}) = \TT{cd}(v_{13}) = 3$. The traversal removal
algorithm starts DFSs from roots $v_{10}$ and $v_{11}$,
respectively. Assume the first DFS starts at $v_{10}$. The fact of
$\TT{cd}(v_{10}) < 3$ implies that $v_{10}$ does not have enough
neighbors in the $3$-core, and therefore $\TT{core}(v_{10})$ needs to be
updated. The algorithm appends $v_{10}$ into $V^*$ and decreases
$\TT{core}(v_{10})$ by $1$.  Due to the change of $\TT{core}(v_{10})$,
both $\TT{cd}(v_{12})$ and $\TT{cd}(v_{13})$ decrease by $1$ and
become $2$.  The algorithm continues DFS from these two vertices
respectively to find more vertices in $V^*$.
% 
% We omit the
% remaining details here because they are almost the same as the case
% for $v_{10}$.
%
Finally,  $V^* = \{v_{10}, v_{11}, v_{12}, v_{13}\}$.
\end{example}
}

It is important to note that $\TT{pcd}$ needs to be maintained for the
following possible edge insertions, even though $\TT{pcd}$ is not
required in the traversal removal algorithm.  The performance of the
algorithm is completely dominated by the maintenance of \TT{pcd}.  By
the definition, $\TT{pcd}(u)$ may be affected by a vertex $w$ that is
$2$ hops away from $u$. If $\TT{core}(u)$ is decreased by $1$ for every
$u \in V^*$, in the worst case, $\mathcal{O}(\sum_{v \in
  \TT{nbr}(V^*)}\TT{deg}(v))$ vertices may have their $\TT{pcd}$
affected.

Both the traversal insertion algorithm and the traversal removal algorithm
suffer from the same performance problem to maintain \TT{pcd}, and the
cost to maintain $\TT{pcd}$ is high.  In some cases, although the
algorithm visits only a small number of vertices to find $V^*$, the
benefit brought however is neutralized by the cost to maintain
\TT{pcd}.

\section{An Order-Based Algorithm} \label{sec:basic-algs}

There are two main issues in the traversal
algorithm.  One is the cost of finding $V^*$ and the other is the cost
of maintaining \TT{pcd}. The issue of finding $V^*$ by the traversal
insertion algorithm is more serious than the traversal removal
algorithm. Unlike the traversal removal algorithm, the traversal
insertion algorithm does not take the idea shown in 
\TT{\CD} (Algorithm \ref{alg:core-decomposition}).
%
% How to search for $V^*$ efficiently is an important task in core
% maintenance.  For the insertion case, instead of using the traversal
% insertion algorithm, alternatively we can find $V^*$ by using the same
% idea as in Algorithm \ref{alg:core-decomposition}.
%
Recall the idea shown in \TT{\CD}. When a new edge $(u, v)$ with $K =
\TT{core}(u) \leq \TT{core}(v)$ is inserted, it identifies $V^*$ by
repeatedly removing vertices whose degrees are less than $K +
1$. Then, $V^*$ is identified as the set of the vertices taken from
the vertices remained if their original core numbers are $K$.
% 
% When there is no more
% vertex to be removed,
%
The main reason that the traversal insertion algorithm does not do so
is due to the fact that the cost of core maintenance otherwise will become 
as high as the cost of core decomposition.

In this work, we revisit the idea presented in \TT{\CD} (Algorithm
\ref{alg:core-decomposition}), and design new algorithms to
significantly improve the performance for core maintenance for both
edge insertions and removals.  The challenges by taking the idea
behind \TT{\CD} for core maintenance are twofold.
First, we need to find an efficient way to maintain the degrees of the
neighbors when a vertex is removed during core maintenance, instead of
removing vertices as done by \TT{\CD}.  In other words, we need to
maintain the order used by \TT{\CD}, but we cannot afford to compute
the order in core maintenance.
Second, because the number of vertices to be affected can
be large, it is impractical to remove vertices online one-by-one.
The solution we propose to address the challenges is to introduce an
order, called $k$-order, with which we can reduce the core maintenance
cost.

Consider a graph $G$. Let $G^0 = G$ and $G^i$ be
the graph after $i$ insertion/removal of edges.
The $k$-order is initially defined over $G^0$, which is the order by
\TT{\CD} (Algorithm \ref{alg:core-decomposition}). In details, the
$k$-order, denoted as $\order$, is that $u \order v$ if and only if
vertex $u$ will be removed in Algorithm
\ref{alg:core-decomposition} before $v$. It is obvious that $u \order
v$ if $\TT{core}(u) < \TT{core}(v)$.  It is worth mentioning that
either $u \order v$ or $v \order u$ is possible when
$\TT{core}(v) = \TT{core}(u)$.
In such a case, one of the two can be
used, if it can be obtained by Algorithm \ref{alg:core-decomposition}.
When $\order$ is determined for $G^0$, we will maintain the order
$\order$ such that it is a $k$-order for graph $G^i$ for $i >
0$.
In the following, for simplicity, we use $G$ instead of $G^i$ and
$G'$ instead of $G^{i+1}$ when the context is clear.

\begin{definition} \label{def:k-order0}
({\bf $k$-order}) Given a graph $G$, let $G^0$ be $G$. Assume $G^i$
  is a graph after $i$ insertion/removal of edges.
%   $G^{i+1}$ is a graph of
%   $G^i$ by either inserting a new edge into $G^i$ or removing an edge
%   from $G^i$.
The $k$-order $\order$ is defined for any
$u$ and $v$ over $G^i$ as follows.
\begin{itemize}[noitemsep,topsep=5pt]
\item When $\TT{core}(u) < \TT{core}(v)$, $u \order v$.
\item When $\TT{core}(u) = \TT{core}(v)$, $u \order v$ 
      if $u$ is removed before $v$ by \TT{\CD} (Algorithm
  \ref{alg:core-decomposition}) for $G^i$.
\end{itemize}
A $k$-order, $(v_1, v_2, \ldots, v_n)$, for $v_i \in V(G)$, is an
instance of all the possible vertex sequences produced by Algorithm
\ref{alg:core-decomposition}.  The transitivity holds; that
is, $u \order v$ if $u \order w$ and $w \order v$.
\end{definition}

\comment{
It is important to note that we will show that we do not need to
recompute $u \order v$ for $G^i$, and we can maintain the same
$k$-order as initially computed for $G^0$.
}

\begin{definition} ({\bf remaining degree})
  \label{def:rem} For a vertex $u$ in $G^i$,
  the remaining degree of $u$, denoted as $\rem(u)$, is defined
  as: $\rem(u) = |\{v \in \TT{nbr}(u): u \order v\}|$.
\end{definition}

Here, $\rem(u)$ is the degree of the neighbors that appear after
$u$.
Given $k$-order $\order$ defined, we use $O_k$ to denote the sequence
of vertices in $k$-order whose core numbers are $k$.  It is obvious
that for a vertex $u$ in $O_k$, $\rem(u) \leq k$.  And we have a
sequence of $O_0O_1O_2 \cdots$, where $O_i \order O_j$ if $i < j$.
It is clear that $\order$ defined over the sequence of
$O_0O_1O_2 \cdots$ is implied by the $k$-order ($\order$).

\begin{lemma} \label{lmm:rem}
Given a graph $G$, the order ($\order$) on $O_0O_1O_2\cdots$ over
$V(G)$ is the $k$-order if and only if $\rem(v) \leq k$
for every vertex $v$ in $O_k$, for $\forall k$. 
\end{lemma}

\proofsketch Please refer to the appendix.

By Lemma~\ref{lmm:rem}, in order to maintain the $k$-order, we
maintain every $O_k$ such that $\forall v \in O_k$, $\rem(v) \leq k$.

\begin{figure}[t]
\centering
\epsfig{file=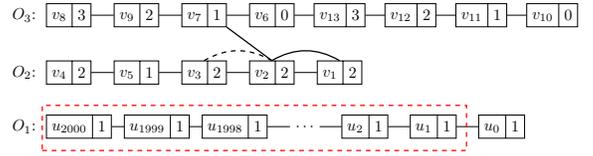, scale=0.6}
\caption{The $k$-order for $G$ in \figurename~\ref{fig:example}} \label{fig:list}
\end{figure}

\begin{example} \label{exm:rem}
In \figurename~\ref{fig:list}, we show a $k$-order for the graph
$G$ (\figurename~\ref{fig:example}) by showing $O_1$, $O_2$, $O_3$. In
$O_1$, $u_i \order u_j$ if $i > j$.  In $O_2$, $v_4 \order v_5 \order v_3
\order v_2 \order v_1$.  In $O_3$, $v_8 \order v_9 \order v_7 \order v_6
\order v_{13} \order v_{12} \order v_{11} \order v_{10}$. The $k$-order is
one of the orders computed by Algorithm \ref{alg:core-decomposition}
for the graph $G$.
Consider the vertex $v_2$ in $G$. $v_2$ has three neighbors: $v_1$,
$v_3$, and $v_7$. Here, both $v_1$ and $v_3$ are in $O_2$ and $v_7$ is
in $O_3$, because $\TT{core}(v_1) = \TT{core}(v_3) = 2$ and
$\TT{core}(v_7) = 3$. In terms of $k$-order, $v_3 \order v_2 \order v_1
\order v_7$. Therefore, $\rem(v_2) = 2$.
The $\rem(v_i)$ for $v_i$ is shown as a number next to $v_i$ in \figurename~\ref{fig:list}.
% 
% Because $v_3 \order v_2$ in $O_2$, $v_3$ is not counted in $\rem(v_2)$.
% However, for $v_1$ and $v_7$, they are involved in $\rem(v_2)$ since $v_2
% \order v_1$ and $\TT{core}(v_7) > \TT{core}(v_2)$.
%
\end{example}

\comment{
Essentially, $O_k$ specifies an order in which vertices in $S_k$ can
be removed during core decomposition. The rationale is as follows. If
we remove vertices in $S_k$ in the order of $O_k$, it is guaranteed
that when we want to process a vertex $v$, its degree at that time is
$\rem(v)$, which is at most $k$. Therefore, the vertex $v$ can not
be in the $(k + 1)$-core, thus can be removed.
}

\begin{figure}[t]
\begin{subfigure}{\columnwidth}
\centering
\epsfig{file=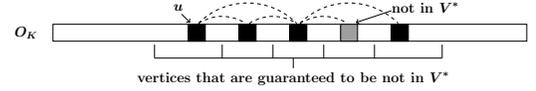, scale=0.45}
\caption{Identification of $V^*$ in the Insertion Algorithm}\label{fig:jump}
\end{subfigure}
\begin{subfigure}{\columnwidth}
\centering
\epsfig{file=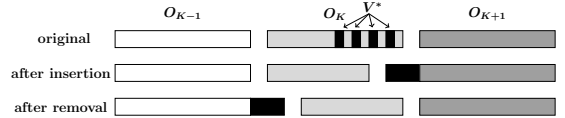, scale=0.45}
\caption{Maintenance of $k$-Orders}\label{fig:maintenance}
\end{subfigure}
\vspace{-0.2cm}
\caption{Algorithm Framework} \label{fig:framework}
\end{figure}

\subsection{An Algorithm Overview}

We design new algorithms to deal with an edge $(u, v)$ to be
inserted into or removed from $G$.  For
simplicity, we assume $K = \TT{core}(u) \leq \TT{core}(v)$, and $u
\order v$. Here, $u \in O_K$.  Let the resulting graph of the
insertion/removal be $G'$.  There are two key issues in our algorithms.
First, we need to identify $V^*$, the set of vertices whose core numbers need to
be updated,
efficiently using the $k$-order $\order$ of $G$. Second, we need to maintain the $k$-order efficiently such that the new
order $\order'$ is the $k$-order of $G'$.

\vspace{-0.15cm}

\stitle{Edge Insertion:} Suppose an edge $(u, v)$ is inserted for $u
\in O_K$ and $u \order v$. $V^*$ can be computed efficiently. We show
the main ideas below.
\vspace{-0.2cm}
\begin{enumerate}[noitemsep,topsep=5pt]
\renewcommand{\labelenumi}{(i\arabic{enumi})}
\item A vertex $w$ cannot be in $V^*$ if $w \in O_{L}$
  for $L < K$.  
\item A vertex $w$ cannot be in $V^*$ if $w \in O_{L}$
  for $K < L$.
\item A vertex $w$ in $O_K$ cannot be in $V^*$ if $w \order u$.
\item A vertex $w$ in $O_K$ \textit{may} be in $V^*$ if $u \order w$ and
  there is a path $w_0, w_1, w_2, \cdots, w_t$ such that
  $w_0 = u$, $w_t = w$, $(w_i, w_{i+1}) \in E$ and $w_i \order w_{i+1}$
  for $0 \leq i < t$.
\item A vertex $w$ that satisfies (i4)
  is not necessarily in $V^*$.
\end{enumerate}

\vspace{-0.2cm}

Items (i1) and (i2) are proved by Theorem \ref{thm:core-region}.
We further explain other points using \figurename~\ref{fig:jump}.
In $O_K$, a vertex $w$ can not be
in $V^*$ if $w \order u$ by (i3). \figurename~\ref{fig:jump} also
shows that the vertices reachable from $u$ by (i4) are possibly in
$V^*$, but not all the vertices reachable from $u$ are included in
$V^*$ by (i5).

To compute $V^*$, we design an efficient algorithm that can ``jump''
from a candidate of $V^*$ to another following the $k$-order of
$G$.
% 
% The ability to jump significantly reduces the number of
% vertices to visit, in contrary to the raw \TT{CoreDecomposition} algorithm
% (Algorithm \ref{alg:core-decomposition}).
%
Therefore, the search space for obtaining $V^*$ is significantly
reduced.
Suppose $V^*$ is obtained.  To obtain the $k$-order $\order'$ of the
resulting graph $G'$, as shown in \figurename~\ref{fig:maintenance}, the set
of vertices in $V^*$ in $O_K$ will be moved to the beginning of
$O_{K+1}$. The $k$-order by the insertion of $(u, v)$ remains
unchanged for all vertices in $O_{L}$, for either $K < L$ or $K > L$,
in $G'$. The $k$-order for the vertices that will be in the same $O_K$
in the resulting graph $G'$ needs small adjustment. It is guaranteed
that only vertices visited by our algorithm may have their positions
changed. The new order $\order'$ is a valid $k$-order of $G'$.

\vspace{-0.25cm}

\stitle{Edge Removal:} Suppose an edge $(u, v)$ is removed.
We adopt the similar idea presented in the traversal removal
algorithm to find $V^*$. However, we maintain the $k$-order
instead of $\TT{pcd}$ used in the traversal removal algorithm
to achieve high efficiency.
%
% The
% maintenance of $k$-orders in the removal case is easy too.
%
To obtain the $k$-order of $G'$, we move the vertices in $V^*$ from
$O_K$ to the end of $O_{K-1}$, without affecting the original vertices
in $O_{K-1}$, as shown in \figurename~\ref{fig:maintenance}.  The resulting
order is guaranteed to be a valid $k$-order of $G'$.  We
present more details when we discuss our removal algorithm.

\vspace{-0.3cm}

\subsection{The Order-Insertion Algorithm}

\vspace{-0.2cm}

\stitle{Identifying $V^*$:}
Suppose an edge $(u, v)$ is inserted for $u \in O_K$ and $u \order v$.
Since (i1) and (i2) are trivial, we focus on (i3) -- (i5).  We call
the vertex $u$ as a root since $u \order v$. Note that by insertion of
$(u, v)$, $\rem(u)$ is increased by $1$ to reflect the insertion
of $(u, v)$, while $\rem(v)$ remains unchanged. Next, we show in
Lemma~\ref{lmm:instant-termination} that no vertices in $O_K$ can
be in $V^*$ if $u$ does not
have sufficient neighbors that appear after $u$ in $k$-order,
i.e., $\rem(u) \leq K$. This implies that we need to update
core numbers of vertices in $O_K$ only if $\rem(u) > K$.  We also show in
Lemma~\ref{lmm:before-root} that there is no need to update the
vertices that appear before $u$ in $O_K$, when $\rem(u) > K$.  For
the vertices that appear after $u$ in $k$-order in $O_K$, we show the
cases we need to consider.
% and give some additional lemma.

\vspace{-0.1cm}

\begin{lemma} \label{lmm:instant-termination}
No core number needs to be updated for vertices in $O_K$, if
$\rem(u) \leq K$ after increasing $\rem(u)$ by $1$.
\end{lemma}

\vspace{-0.1cm}

\begin{lemma} \label{lmm:before-root}
No vertex $w$ that appears before $u$ in $O_K$ can be
in $V^*$, if $\rem(u) > K$ after increasing $\rem(u)$ by $1$.
\end{lemma} 

\begin{proofsketch}
The proof of Lemma~\ref{lmm:instant-termination} and
\ref{lmm:before-root} is in the appendix.
\end{proofsketch}

\begin{example} \label{exm:klist}
Reconsider Example \ref{exm:traversal-ins}. Suppose edge $(v_4,
u_0)$ is inserted to $G$, where $u_0 \order v_4$ in $G$.  Then,
$\rem(u_0)$ becomes $2$.
Note that $u_0$ is in $O_1$ before the
new edge is inserted. By Lemma~\ref{lmm:before-root}, all vertices
that occur before $u_0$ in $O_1$ (the rectangle in
\figurename~\ref{fig:list}) can not be in $V^*$.  Reconsider $u_0$,
$\rem(u_0) = 2$ implies that $u_0$ has at least two neighbors
appearing after $u_0$ in the $k$-order, which are $v_4$ and $v_5$ in this
example.
Hence, $u_0$ will be in the $2$-core of the new graph $G
\cup \{(v_4, u_0)\}$. We have $V^* = \{u_0\}$. As shown in this
example, our approach achieves high efficiency, because we only
need to visit $1$ vertex (i.e., $u_0$), while the traversal insertion
algorithm needs to visit $1,999$ vertices in total to identify
$V^*$.
\end{example}

In Example~\ref{exm:klist}, vertex $u_0$ is the last vertex in
$O_1$. Generally, when a new edge $(u, v)$ is inserted
for $u \in O_K$ and $u \order v$, $u$ can be in any position in
$O_K$. There may be vertices appearing after $u$ in
$O_K$. Among all such vertices, we further discuss whether a vertex
$w$ with $u \order w$ in $O_K$ can be a potential candidate in $V^*$.
We first introduce the concept of \extdegree.

\begin{definition}  ({\bf \extdegree})
  \label{def:ext} For a vertex $w$ in $O_K$,
  the \extdegree of $w$, denoted as $\ext(w)$, is defined
  as: $\ext(w)$ $= |\{w' \in \TT{nbr}(w): w' \order w \land
  \TT{core}(w') = K
  \land w'
  \NM{ is a potential candidate of $V^*$}\}|$.

\end{definition}

In order to test whether a vertex $w$ can be a potential candidate in
$V^*$, we use $\ext(w)+\rem(w)$.  Recall that $\rem(w)$
(Definition~\ref{def:rem}) counts the number of $w$'s neighbors after $w$
in the $k$-order, whose core numbers are greater than or equal to $K$.
In other words, $\rem(w)$ counts how many $w$'s neighbors after $w$ can
be in the new $(K + 1)$-core.
Therefore, $\rem(w) + \ext(w)$ upper bounds the number of $w$'s
neighbors in the new $(K + 1)$-core. Specifically,
if $\rem(w) + \ext(w) \leq K$, $w$ does not have sufficient neighbors
in the new $(K + 1)$-core, thus $w$ cannot be in $V^*$; if $\rem(w) + \ext(w) >
K$ otherwise, $w$ is a potential candidate.

Initially, $\ext(w) = 0$ for each $w \in O_K$ since we have not found
any candidate yet.
We start from $u$ and visit vertices in $O_K$ following the $k$-order.
For each vertex $w$ being visited, we discuss two cases.

\stitle{Case-1} ($\ext(w) + \rem(w) > K$): Since $w$ potentially has
more than $K$ neighbors in the new $(K + 1)$-core, $w$ may be in
$V^*$, i.e., $\TT{core}(w)$ may become $K + 1$. We put $w$ in a set
$V_C$, which records all current visited potential
candidates. Further, each neighbor $w'$ of $w$ that $w \order w'$ in
$O_K$, obtains one \extdegree ($\ext(w')$ increased by $1$) to
reflect the existence of $w$.  In other words, for each such $w'$, the
potential number of neighbors in the new $(K + 1)$-core increases by
$1$.  Note that $u$ is a special example in this case because
$\ext(u) = 0$ and $\rem(u) > K$. We next visit the vertex that is
next to $w$ in the $k$-order.

\stitle{Case-2} ($\ext(w) + \rem(w) \leq K$): $w$ does not have
sufficient neighbors in the new $(K + 1)$-core, since $\ext(w) + \rem(w) \leq
K$. Therefore, $w \notin V^*$.
There are two subcases, namely, {\bf Case-2a} for $\ext(w) = 0$ and {\bf Case-2b} for
$\ext(w) \neq 0$.

For {\bf Case-2a}, $w$ cannot be in $V^*$. We next visit the vertex
$w''$ which is next to $w$ in the $k$-order.  It is possible that
$w''$ again is in {\bf Case-2a}, implying that $w'' \notin V^*$. More
generally, let $w'$ be the first vertex after $w$ in $O_K$ such that
$\ext(w') \neq 0$. For any $w''$ in the range $[w, w')$ in the
  $k$-order, $\ext(w'') = 0$. As a result, these $w''$s cannot be in
  $V^*$ and we can skip them, directly ``jump'' to visit $w'$
  next. Specially, if such $w'$ does not exist, all vertices after $w$
  cannot be in $V^*$.

% ***********************************************************************

\begin{figure}[t]
\centering
\epsfig{file=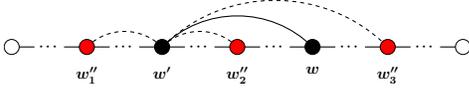,scale=0.5}
\caption{Demonstration of {\bf Case-2b}.}
\label{fig:case2b}
\end{figure}

For {\bf Case-2b}, it differs from {\bf Case-2a} for $\ext(w)
\neq 0$. 
%
% $w \not \in V^*$. 
%
Here, $\ext(w) \neq 0$ implies that some neighbor $w'$ that $w'
\order w$ is considered as a potential candidate. In other
words, $\ext(w') + \rem(w') > K$ (refer to {\bf Case-1} where $w'$ is
in the position of $w$ in {\bf Case-1}).
Excluding $w$ from $V^*$ results in $\rem(w')$ to be decreased by $1$ for each
neighbor vertex $w'$ that $w' \order w$ and $w' \in V_C$.
Recall that $V_C$ is the set of vertices that currently are candidates for
$V^*$.
If $\ext(w') +
\rem(w') \leq K$ after such a update, then
we can remove $w'$ from $V_C$ since $w'$ is not a candidate any more,
which may further result in other potential candidates to be removed
from $V_C$.
The situation can be demonstrated using \figurename~\ref{fig:case2b}. Suppose
$w'$ is not a potential candidate after $\rem(w')$ decreases by 1.  Then (1)
$\rem(w''_1)$ is decreased by $1$ for each neighbor $w''_1$ of $w'$ that
$w''_1 \in V_C$ and $w''_1 \order w'$; (2) $\ext(w''_2)$ is decreased
by $1$ for each neighbor $w''_2$ of $w'$ that $w''_2 \in V_C$ and $w'
\order w''_2 \order w$; (3) $\ext(w''_3)$ is decreased by $1$ for each
neighbor $w''_3$ of $w'$ that $w \order w''_3$.  For $w''_1$ and
$w''_2$, they may not be a candidate any more after update and will be
removed from $V_C$. The chain effect can further propagate. For
$w''_3$, we just update its $\ext$ and it will be processed when we
visit it later. Recall that we visit vertices in $O_K$ following the
$k$-order.  We choose the vertex next to $w$ in the $k$-order as the
next vertex to visit.

When we reach the end of $O_K$, $V^* = V_C$. We explain the rationale
as follows. For each $w \in V_C$, we have
(1) $\ext(w) = |\{w' \in \TT{nbr}(w) : w' \in V_C \land w' \order w\}|$, and
(2) $\rem(w) = |\{w' \in \TT{nbr}(w) : w' \in V_C \land w \order w' \NM{ or }
\TT{core}(w') > K\}|$.
Therefore, $\ext(w) + \rem(w)$ is the number of the neighbors of $w$
that are in $V_{> K} \cup V_C$ and is greater than or equal to $K +
1$, where $V_{>K} = \{v \in V : \TT{core}(v) > K\}$. Hence, $G(V_{>K}
\cup V_C)$ is the new $(K + 1)$-core of $G'$. In
other words, $V^* = V_C$.
%
% Therefore, $\ext(w) + \rem(w) = \TT{deg}(w, V_{> K} \cup V_C) \geq K +
% 1$, where $V_{>K} = \{v \in V : \TT{core}(v) > K\}$. We conclude that
% $V_{>K} \cup V_C$ is the new $(K + 1)$-core of $G'$. In other words,
% $V^* = V_C$.
%

% 
% We next consider the case when the vertex being visited has non-zero extra
% degrees. Consider $v$ in \figurename~\ref{fig:after-root}, whose extra degree
% $\ext(v) > 0$ ($\ext(v) = 1$).
% In general, there are two cases to consider:
% 
% \begin{itemize}
%   \item $\rem(v) + \ext(v) \leq K$. $\ext(v)$ counts the number of
%   $v$'s neighbors that are before $v$ in $O_K$ and have not been removed yet.
%   Therefore, $\rem(v) + \ext(v)$ actually is the maximum possible number
%   of $v$'s neighbors in the $(K + 1)$-core of $G \cup \{(u_1, u_2)\}$. If
%   $\rem(v) + \ext(v) \leq K$, $v$ can be safely removed. The removal of
%   $v$ may have chain effect. The rationale is as follows: After $v$ is removed,
%   for each neighbor $u$ of $v$ that $u \order v$ and has not been removed yet,
%   $\rem(u)$ decreases by $1$. As a result, $\rem(u) + \ext(u)$ may
%   become smaller than or equal to $K$ and then $u$ can be removed. We will show
%   more details when we present the basic insertion algorithm in the next
%   subsection.
%   \item $\rem(v) + \ext(v) > K$. In this case, $\TT{core}(v)$ may
%   need to be updated and thus $v$ can not be removed. For each neighbor $u$ of
%   $v$ that $v \order u$, $u$ obtains one more extra degree.
% \end{itemize}

\stitle{Maintaining $k$-order:} Since the core numbers of vertices in
$V^*$ change from $K$ to $K+1$, we need to obtain two new sequences
$O'_K$ and $O'_{K+1}$ for vertices with core number $K$ and $K + 1$
respectively, such that the new order $\order'$ over $O_0\cdots
O_{K-1}O'_KO'_{K+1}O_{K+2}\cdots$ is a $k$-order of $G'$.
By Lemma~\ref{lmm:rem}, we only need to guarantee that (1) $\rem(w) \leq K$ for
$\forall w \in O'_K$ and (2) $\rem(w) \leq K + 1$ for $\forall w \in O'_{K+1}$.
We discuss $O'_K$ first.

In order to construct $O'_K$, the main idea is to append a vertex $w$
to $O'_K$ when we encounter a vertex $w$ with $\ext(w) + \rem(w) \leq
K$ during the process of identifying $V^*$. The rationale is that
$\ext(w) + \rem(w)$ is exactly the number of $w$'s neighbors that have
not been appended to $O'_K$ yet or that have core number greater than
$K$. Therefore, $\ext(w) + \rem(w)$ is the number of $w$'s neighbors
that will appear after $w$ in the new $k$-order.  The new $\rem(w)$ is
set as $\ext(w) + \rem(w)$. We give the details.  Let $w$ be the
vertex currently being visited.
\begin{itemize}[noitemsep,topsep=5pt]
\item {\bf Case-1}: $w$ cannot be appended to $O'_K$
  temporarily and thus is skipped.
\item {\bf Case-2a}: $w$ is appended to $O'_K$. Because
  $\ext(w) = 0$, $\rem(w)$ remains unchanged.
\item {\bf Case-2b}: $w$ is appended to $O'_K$ and $\rem(w)$ becomes
  $\ext(w)$ $+ \rem(w)$. For each $w' \in V_C$ that is found
  $\ext(w') + \rem(w') \leq K$, we append $w'$ to $O'_K$ and the new $\rem(w')$
  is set as $\ext(w') + \rem(w')$.
\end{itemize}

In order to construct $O'_{K+1}$, we insert vertices in $V_C$ to the
beginning of $O_{K+1}$ such that $\forall w_1, w_2 \in V_C$, $w_1 \order' w_2$
in the new order $\order'$ if $w_1 \order w_2$ in the original order $\order$.
In this way, the new
$\rem(w)$ for each $w \in V_C$ is not greater than the original $\rem(w)$.
Therefore, the new $\rem(w)$ is smaller than or equal to $K + 1$. Note that we
do not need to update the $\rem$ for vertices in the original $O_{K+1}$. This
explains why we insert $V_C$ to the beginning of $O_{K+1}$ rather than other
positions.

\stitle{The Algorithm:}  The pseudocode of the insertion
algorithm is shown in Algorithm \ref{alg:basic-insert}. While the
algorithm is searching for $V^*$, it maintains the $k$-order at the
same time. It consists of three phases: a) the preparing phase
(lines~\ref{line:basic-insert-prepare-beg}-\ref{line:basic-insert-prepare-end});
b) the core phase
(lines~\ref{line:basic-insert-core-beg}-\ref{line:basic-insert-core-end});
c) the ending phase
(lines~\ref{line:basic-insert-post-beg}-\ref{line:basic-insert-post-end}).

\begin{algorithm}[t]
{\small
\linespread{0.8}\selectfont
\SetKwInOut{Input}{input} \SetKwInOut{Output}{output} \SetKw{Each}{each}
\SetKw{SuchThat}{such that}
\Input{$G = (V, E)$: the input graph; $(u, v)$: the edge to insert}

\tcc{Preparing Phase}
$K \leftarrow \min\{\TT{core}(u), \TT{core}(v)\}$\;
\label{line:basic-insert-prepare-beg}
let the vertices in $O_K$ be $v_1, v_2, \cdots, v_{|O_K|}$ in
$k$-order\;
%
% $\TT{rank}(v_i) \leftarrow i$ for $\forall v_i$ \NM{ in } $O_K$\;
% \label{line:basic-insert-rank}
%
$G' \leftarrow G \cup \{(u, v)\}$\;
% $r \leftarrow u$ assuming $u \order v$\;
% 
% \If{$(\TT{core}(u_1) = \TT{core}(u_2) \wedge \TT{rank}(u_2) < \TT{rank}(u_1))
%      \linebreak \NM { or } (\TT{core}(u_2) < \TT{core}(u_1))$}
% {
% 	$r \leftarrow u_2$\;
% }
%
$\rem(u) \leftarrow \rem(u) + 1$ assuming $u \order v$\;
\label{line:basic-insert-u-incr}
\label{line:basic-insert-prepare-end}

\tcc{Core Phase}

{
\lIf {$\rem(u) \leq K$} {
	\label{line:basic-insert-core-beg}
	\label{line:basic-insert-early-termination}
    \Return;
}
}

$O'_K \leftarrow \emptyset$; $V_C \leftarrow \emptyset$\;
$i \leftarrow 1$\;
\While {$i \leq |O_K|$}
{
	\label{line:basic-insert-for-beg}
	\uIf{$\ext(v_i) + \rem(v_i) > K$}
	{
	  \label{line:basic-insert-case1}
                remove $v_i$ from $O_K$ and
                append it to $V_C$\;
%		$\TT{keep}(v_i) \leftarrow \TT{true}$\;
		\For {\Each $w \in \TT{nbr}(v_i)$ \SuchThat
			  $\TT{core}(w) = K \land v_i \order w$}
		{
			$\ext(w) \leftarrow \ext(w) + 1$\;
			\label{line:basic-insert-case1-incr}
		}
		$i \leftarrow i + 1$\;	
		\label{line:basic-insert-case1-end}

	}
	\uElseIf{$\ext(v_i) = 0$}
	{
		\label{line:basic-insert-case2a}
		let $v_j \in O_K$ be the first vertex for $v_i
                \order v_j$ such that $\ext(v_j) > 0$ or
		$\rem(v_j) > K$\; \label{line:basic-insert-j}
		\lIf{\NM{such }$v_j$\NM{ does not exist}} {$j \leftarrow |O_K| + 1$;}
		
		remove $v_i, \cdots, v_{j-1}$ from $O_K$\;
		\label{line:basic-insert-remove}
		append $v_i, \cdots, v_{j-1}$ to $O'_K$ in order\;
		\label{line:basic-insert-insert}
		$i \leftarrow j$\;
		\label{line:basic-insert-case2a-end}	
	
	}
	\Else
	{
	
		\label{line:basic-insert-case2b}
		remove $v_i$ from $O_K$ and append to $O'_K$\;
		$\rem(v_i) \leftarrow \rem(v_i) + \ext(v_i)$\;
		$\ext(v_i) \leftarrow 0$\;
		\TT{RemoveCandidates}($G'$, $V_C$, $O'_K$, $v_i$, $K$)\;
		\label{line:basic-insert-call}
		$i \leftarrow i + 1$\;
		\label{line:basic-insert-case2b-end}
		\label{line:basic-insert-core-end}
		\label{line:basic-insert-for-end}
	}
}
\tcc{Ending Phase}

$V^* \leftarrow V_C$\;
\label{line:basic-insert-post-beg}
\For{\Each $w \in V^*$}
{
	$\ext(w) \leftarrow 0$\; %\quad
%	$\TT{keep}(w) \leftarrow \texttt{false}$\;
	$\TT{core}(w) \leftarrow \TT{core}(w) + 1$\;
}
insert vertices in $V^*$ to the beginning of $O_{K + 1}$ in
$k$-order\;
let the $k$-order for $O_K$ in $G'$ be $O'_K$\;
$G \leftarrow G'$\;
% $O_K \leftarrow O$\;
update $\TT{mcd}$ accordingly\; \label{line:basic-insert-post-end}
}
\caption{\TT{\OI}($G$, $(u, v)$)}\label{alg:basic-insert}
\end{algorithm}

\begin{algorithm}[t]
{\small  
\linespread{0.8}\selectfont
\SetKwInOut{Input}{input} \SetKwInOut{Output}{output}
\SetKw{SuchThat}{such that} \SetKw{Each}{each} \SetKw{Not}{not}
% \Input{$G = (V, E)$: the input graph; $O$, $v$, $K$:
% please refer to Algorithm \ref{alg:basic-insert}}
$\mathcal{Q} \leftarrow $ an empty queue\;
\label{line:keep-init-beg}
\For{\Each $w' \in \TT{nbr}(w)$ \SuchThat $w' \in V_C$}
{
	$\rem(w') \leftarrow \rem(w') - 1$\;
	\If{$\rem(w') + \ext(w') \leq K$}{
%		$\TT{visited}(w) \leftarrow \TT{true}$\;
		$\mathcal{Q}$.enqueue($w'$)\;
		\label{line:keep-init-end}
	}
}
%\While{\Not $\mathcal{Q}.\NM{empty}()$}
\While{$\mathcal{Q} \neq \emptyset$}
{
	\label{line:keep-while-beg}
	$w' \leftarrow \mathcal{Q}$.dequeue()\;
%	$\TT{visited}(w') \leftarrow \TT{false}$\;
        %	$\TT{keep}(z) \leftarrow \TT{false}$\;
	$\rem(w') \leftarrow \rem(w') + \ext(w')$;
	\quad $\ext(w') \leftarrow 0$\;
        remove $w'$ from $V_C$; \quad append $w'$ to $O'_K$\;
	%{\DontPrintSemicolon \; \label{line:keep-new-code1}}
	\For{\Each $w'' \in \TT{nbr}(w')$ \SuchThat $\TT{core}(w'') = K$}
	{
		\label{line:keep-for}
		\uIf {$w \order w''$}
		{
			\label{line:keep-case1}
			$\ext(w'') \leftarrow \ext(w'') - 1$\;
			\label{line:keep-ext-desc}
			%{\DontPrintSemicolon \; \label{line:keep-new-code2}}
		}
		\uElseIf {$w' \order w'' \land w'' \in V_C$}
		{
			\label{line:keep-case2}
			$\ext(w'') \leftarrow \ext(w'') - 1$\;
			\If {$\ext(w'') + \rem(w'') \leq K
                          \wedge w'' \not \in {\cal Q}$}
			{
				\label{line:keep-tmp1}
%				$\TT{visited}(w'') \leftarrow \TT{true}$\;
				$\mathcal{Q}$.enqueue($w''$)\;
				\label{line:keep-tmp2}
			}
		}
		\ElseIf{$w'' \in V_C$}
		{
			\label{line:keep-case3}
			$\rem(w'') \leftarrow \rem(w'') - 1$\;
			same codes as lines \ref{line:keep-tmp1} to \ref{line:keep-tmp2}\;
			\label{line:keep-while-end}
		}
	}
}
}
\caption{RemoveCandidates($G'$, $V_C$, $O'_K$, $w$, $K$)} \label{alg:keep}
\end{algorithm}

In the preparing phase, we set $K$ as the smaller one of $\TT{core}(u)$
and $\TT{core}(v)$. By Theorem \ref{thm:core-region}, only vertices in
$O_K$ may need to have core numbers updated.
The new edge $(u, v)$ then is inserted to $G$, and
$\rem(u)$ is increased by $1$ to reflect the insertion of $(u, v)$,
assuming $u \order v$.

In the core phase, if $\rem(u) \leq K$, $O_K$ is still valid and no
core number needs to be updated (Lemma
\ref{lmm:instant-termination}). We can terminate the algorithm
(line~\ref{line:basic-insert-early-termination}).  On the other hand,
if $\rem(u) > K$, we deal with the 3 cases using a while loop
(lines~\ref{line:basic-insert-for-beg}-\ref{line:basic-insert-for-end}),
where each iteration contains three conditional branches,
corresponding to {\bf Case-1} (lines~10-13), {\bf Case-2a}
(lines~15-19) and {\bf Case-2b} (lines~21-25) discussed above. For
{\bf Case-2b},
% (line-\ref{line:basic-insert-case2a}),
we call Algorithm \ref{alg:keep} (line~24)
% (line~\ref{line:basic-insert-call})
to find all vertices in $V_C$ that are not candidates any more.

In the ending phase, $V_C$ contains all and only vertices in $V^*$.
$\TT{core}$ thus is increased by $1$ for each vertex in $V_C$.
Following that, vertices in $V_C$ are inserted to the beginning of
$O_{K + 1}$ as discussed above. $O'_K$ then becomes the $O_K$ of $G'$.
We update $\TT{mcd}$ for all relevant vertices in line
\ref{line:basic-insert-post-end} for edge removals in the future.

\stitle{Correctness of the Algorithm:}
We show the correctness of Algorithm~\ref{alg:basic-insert} in the
following theorem.
% \begin{proofsketch}
The proof can be obtained from the discussion above and is omitted here.
% \eop
% \end{proofsketch}

\begin{theorem} \label{thm:basic-insert-core}
Algorithms \ref{alg:basic-insert} and \ref{alg:keep} correctly update
$\TT{core}$ and correctly maintain $O_K$ and $O_{K+1}$.
\end{theorem}

\stitle{Complexity Analysis:} To analyze the time complexity of
Algorithm \ref{alg:basic-insert}, we make the following assumptions
for the data structures used and we will provide more details about
the implementation later.
\vspace{-0.3cm}
\begin{itemize}[noitemsep,topsep=5pt]
  \item For each $k$, we associate $O_k$ with a data structure
  $\mathcal{A}_k$. For $\forall\ u, v \in O_k$, we can test whether
  $u \order v$ using $\mathcal{A}_k$ in $\mathcal{O}(\log |O_k|)$ 
  time. Since we need to move vertices from $O_K$ to $O_{K+1}$, $\mathcal{A}_k$
  supports insertion (resp. removal) of a single vertex to (resp. from)
  $\mathcal{A}_k$ in $\mathcal{O}(\log |O_k|)$ time.
  \item A data structure $\mathcal{B}$ records all vertices $v_j \in
  O_K$ that $v_i \order v_j$ and $(\ext(v_j) > 0 \lor \rem(v_j) > K)$.
   $\mathcal{B}$ supports ``jumping'' by finding $v_j$ (line
  \ref{line:basic-insert-j} of Algorithm \ref{alg:basic-insert}) in
  $\mathcal{O}(1)$ time. $\mathcal{B}$ supports insertion of
  a single vertex in $\mathcal{O}(\log |O_K|)$ time such that once
  $\ext(w)$ in line
  \ref{line:basic-insert-case1-incr} becomes non-zero or $\rem(u)$ in line
  \ref{line:basic-insert-u-incr} becomes greater than $K$, we can insert $w$ and
  $u$ to $\mathcal{B}$ efficiently.
  $\mathcal{B}$ supports removal of a single vertex in $\mathcal{O}(\log |O_K|)$
  time such that in line \ref{line:keep-ext-desc} of Algorithm \ref{alg:keep},
  we can remove $w''$ from $\mathcal{B}$ efficiently once $\ext(w'')$ is
  decreased to $0$, or remove $v_i$ from $\mathcal{B}$ after $v_i$ is
  processed in Algorithm \ref{alg:basic-insert}.
  \item We assume moving vertices between $O_K$ and $O'_{K}$ or
    between $O_K$ and $O_{K+1}$ is done in $\mathcal{O}(1)$ time.
\end{itemize}

\vspace{-0.3cm}

We denote the set of vertices in {\bf Case-1} and {\bf Case-2b} as
$V^+$. In other words, $V^+$ consists of vertices that enter the first
branch
(line~\ref{line:basic-insert-case1}
in Algorithm~\ref{alg:basic-insert}) or the third branch
(line~\ref{line:basic-insert-case2b}).
Intuitively, this is the set of vertices that we need to visit.
The vertices in {\bf Case-2a} are not considered because
we do not need to expand them, i.e., we do not need to access their
adjacent edges.
%
% thus processing these vertices is cheap.
%
In addition, for {\bf Case-2a}, the algorithm enters the second branch
(line \ref{line:basic-insert-case2a}) at most $|V^+| + 1$ times,
because each time the algorithm enters the second branch, it will
either enter the first or the third branch or terminate in the next
iteration.
We emphasize that $V^* \subseteq V^+$.  We have the following theorem
for the complexity of the algorithm.

\begin{theorem} \label{thm:insertion-complexity}
The time complexity of the insertion algorithm is $\mathcal{O}(\sum_{v \in
V^+}\TT{deg}(v) \cdot \log \max\{|O_K|, |O_{K+1}|\})$.
\end{theorem}

\begin{proofsketch}
Please refer to the appendix.
\end{proofsketch}

\stitle{The size of $V^+$:} We have the following two observations for
$V^+$. First, if a vertex $v$ enters the third branch, our algorithm
in this case will decrease $\rem$ and $\ext$ only for vertices that
have been in $V^+$.  Therefore, this branch does not introduce new
candidates to $V^+$. Second, if a vertex $v$ enters the first branch,
it will increase $\ext$ of its neighbors that are after it in the
$k$-order. These neighbors may be visited and expanded (thus
belong to $V^+$) later by the algorithm. Based on this observation, we
introduce the notion \textit{order core} below.

\begin{definition} \label{def:ss} ({\bf order core})
For each $u \in V(G)$, its order-core, denoted as $\TT{oc}(u)$, is
recursively defined as follows.
{
\setlength{\abovedisplayskip}{3pt}
\setlength{\belowdisplayskip}{3pt}
$$\TT{oc}(u) = \{u\} \cup (\bigcup\textstyle{_{w \in \TT{nbr}(u)\ \wedge\
\TT{core}(u) = \TT{core}(w) \ \wedge u \order w}} \TT{oc}(w))$$
\vspace{-0.5cm}
}
\end{definition}

Essentially, for each $w \in \TT{oc}(u)$, there is a path $w_0$,
$w_1$, $\cdots$, $w_t$ such that $w_0 = u$, $w_t = w$, $(w_i, w_{i+1})
\in E$ and $w_i \order w_{i+1}$ for $0 \leq i < t$.  The lemma below relates
$V^+$, $\TT{oc}(u)$ together.

\begin{lemma} \label{lmm:worst}
Suppose $(u, v)$ is the newly inserted edge. (1) If $\TT{core}(u) <
\TT{core}(v)$, $V^+ \subseteq \TT{oc}(u)$. (2) If
$\TT{core}(u) = \TT{core}(v)$, $V^+ \subseteq \ \TT{oc}(u) \cup \TT{oc}(v)$.
\end{lemma}

\comment{
\begin{table}[t]
{
\resizebox{\columnwidth}{!}{%
\begin{tabular}{|c|l|}
\hline
$\TT{oc}$ & The max \# of vertices that our insertion algorithm visits and
expands \\ \hline 
$\TT{pc}$ & The max \# of vertices that the traversal insertion algorithm visits
and expands \\ \hline 
$\TT{sc}$ & The max \# of vertices that need to visit and expand by Theorem
\ref{thm:core-region} \\ \hline
\end{tabular}
}
}
\caption{Relationship Between $\TT{oc}$, $\TT{pc}$ and $\TT{sc}$.}
\label{tbl:relationship}
\end{table}
}

By Lemma \ref{lmm:worst}, the size of order cores upper bounds the
maximum possible size of $V^+$.
% 
% We summarize $\TT{oc}$, $\TT{pc}$
% (Definition \ref{def:purecore}), $\TT{sc}$ (Definition
% \ref{def:subcore}) in Table \ref{tbl:relationship}.
%
Recall that $\TT{oc}$ is the max \# of vertices that our insertion
algorithm visits and expands, $\TT{pc}$ (Definition~\ref{def:purecore}) is the
max \# of vertices that the traversal insertion algorithm visits and expands,
and $\TT{sc}$ (Section~\ref{sec:background}) is
the max \# of vertices that need to visit and expand by
Theorem~\ref{thm:core-region}. 
We show the cumulative distribution of $\TT{oc}$ for \TT{Patents} and \TT{Orkut}
in \figurename~\ref{fig:size_dist}.
From \figurename~\ref{fig:size_dist}, $\TT{oc}$ has much smaller variation in
the size than $\TT{sc}$ and $\TT{pc}$. While 90\% of vertices have their
$\TT{oc}$ in the order of hundreds or less, 90\% vertices have their
$\TT{pc}$ and $\TT{sc}$ in the order of 10,000 or less. In other words, our
algorithm visits much less vertices than the traversal insertion algorithm
does in the worst case.

In order to investigate the practical size of $V^+$, for the 11 real graphs
tested (Table~\ref{tbl:dataset}), we calculate the ratio {\small$\frac{\NM{sum of }
  |V^+|\NM{ over all insertions}}{\NM{sum of }|V^*|\NM{ over all
    insertions}}$} for our algorithm after inserting 100,000 edges one by one to
each of the graphs. The results are shown in
\figurename~\ref{fig:perf-expansion}. As can be seen, for all graphs,
the ratios are smaller than $4$, which indicates the efficiency of our
algorithm. In addition, we show the distribution of $|V^+|$ in
\figurename~\ref{fig:dist}. The distributions show that the proportion that
$|V^+| > 100$ is negligible for all the 11 graphs, which indicates
that our algorithm is efficient and of small performance variation.

\begin{algorithm}[t]
{\small  
\linespread{0.8}\selectfont
\SetKwInOut{Input}{input} \SetKwInOut{Output}{output}
\SetKw{SuchThat}{such that} \SetKw{Each}{each}
\SetKw{Not}{not}
\Input{$G = (V, E)$: the input graph; $(u, v)$: the edge to remove}
$K \leftarrow \min\{\TT{core}(u), \TT{core}(v)\}$\;
$G \leftarrow G \setminus (u, v)$\;

\lIf{$\TT{core}(u) \leq \TT{core}(v)$}{$\TT{mcd}(u) \leftarrow \TT{mcd}(u) - 1$;}

\lIf{$\TT{core}(u) \geq \TT{core}(v)$}{$\TT{mcd}(v) \leftarrow \TT{mcd}(v) - 1$;}

find $V^*$ using the routine used in the traversal removal algorithm (described
in Section \ref{sec:traversal-remove}) and assume $\TT{core}$ has been correctly updated\;
\label{line:basic-remove-traversal}
\tcc{Update the $k$-order below}
\label{line:basic-remove-maintain-beg}
\For{\Each $w \in V^*$ in the order they are inserted to $V^*$}
{
	$\rem(w) \leftarrow 0$\;
	\For{\Each $w' \in \TT{nbr}(w)$}
	{
		\If{$\TT{core}(w') = K \wedge w' \order w$}
		{
			\label{line:basic-remove-first-if-beg}
			$\rem(w') \leftarrow \rem(w') - 1$;
			\label{line:basic-remove-first-if-end}
		}
		\If{$\TT{core}(w') \geq K \vee w' \in V^*$} {
			\label{line:basic-remove-second-if-beg}
			$\rem(w) \leftarrow \rem(w) + 1$\;
			\label{line:basic-remove-second-if-end}
		}
	}
	$V^* \leftarrow V^* \setminus \{w\}$\;
	\label{line:basic-remove-post-beg}
	remove $w$ from $O_K$ and append it to $O_{K-1}$\;
	\label{line:basic-remove-post-end}
	\label{line:basic-remove-maintain-end}
}
update $\TT{mcd}$ accordingly\; \label{line:basic-remove-mcd}
}
\caption{\TT{\OR}($G$, $(u, v)$)} \label{alg:basic-removal}
\end{algorithm}

\subsection{The Order-Removal Algorithm}

The removal algorithm is presented in Algorithm
\ref{alg:basic-removal}.  First, we remove $(u, v)$ from $G$ and
update $\TT{mcd}$ to reflect the removal of $(u, v)$. The algorithm
adopts the same routine used in the traversal removal algorithm
(Section \ref{sec:traversal-remove}) to find $V^*$.  We maintain the
$k$-order as follows. For each $w \in V^*$ in the order they are
inserted to $V^*$, we update the $\rem$ of $w$ and its neighbors,
remove $w$ from $O_K$, and insert $w$ to the end of $O_{K-1}$. This is
different from the insertion case, where we insert vertices to the
beginning of $O_{K+1}$. Finally, we update $\TT{mcd}$
(line-\ref{line:basic-remove-mcd}) to reflect the change of
$\TT{core}$.

Algorithm \ref{alg:basic-removal} correctly updates $\TT{core}$
because we adopt the same routine used in the traversal removal
algorithm to find $V^*$. For $O_K$ and $O_{K-1}$, we have the
following theorem.

\begin{theorem} \label{thm:basic-remove-list}
Algorithm \ref{alg:basic-removal} correctly maintains $O_K$ and $O_{K-1}$.
\end{theorem}

\vspace{-0.3cm}

\begin{proofsketch}
Please refer to the appendix.
\end{proofsketch}

We adopt the similar idea used in the traversal removal algorithm,
since it requires only $\mathcal{O}(\sum_{v \in
  V^*}\TT{deg}(v))$ to find $V^*$, as discussed in Section
\ref{sec:traversal-remove}.
On the other hand, the critical difference between our \TT{\OR}
(Algorithm \ref{alg:basic-removal}) and the traversal removal
algorithm is the index under maintenance. In the traversal removal
algorithm, it needs to maintain the $\TT{pcd}$ values of the vertices,
while in our algorithm, we instead maintain the $k$-order ($O_K$ and
$O_{K-1}$). As shown in Section \ref{sec:traversal-remove}, the cost
of maintaining $\TT{pcd}$ is usually large and hurts the total
performance significantly.  Different from $\TT{pcd}$, $k$-orders can be
maintained much more efficiently.  Specifically, maintaining the
$k$-order requires only $\mathcal{O}(\log |O_K| \cdot \sum_{w \in
  V^*}\TT{deg}_K(w) + |V^*|\cdot \log |O_{K-1}|)$ time in the worst case,
where $\TT{deg}_K(w)$ counts the number of neighbors $w'$ of $w$ that
$\TT{core}(w') = K$ and is usually small.
Formally, we have the following theorem for the
complexity of the removal algorithm.

\begin{theorem} \label{thm:removal-complexity}
Algorithm \ref{alg:basic-removal} runs in
$
\mathcal{O}(\sum_{w \in V^*} \TT{deg}(w) + \log |O_K|\cdot \sum_{w \in V^*}
\TT{deg}_K(w) + |V^*|\cdot \log |O_{K-1}|) $
time.
\end{theorem}

\vspace{-0.1cm}
\begin{proofsketch}
Please refer to the appendix.
\end{proofsketch}

\vspace{-0.1cm}
\section{Implementation} \label{sec:list}

\stitle{Generation of the $k$-Order:} We generate the $k$-order for a
graph $G$ based on \TT{\CD} (Algorithm \ref{alg:core-decomposition})
by inserting the following code immediately after line \ref{line:k-order-gen} of
Algorithm \ref{alg:core-decomposition}.

\vspace*{-0.2cm}
\noindent \rule{\columnwidth}{1pt}

\begin{figure}
\centering
\epsfig{file=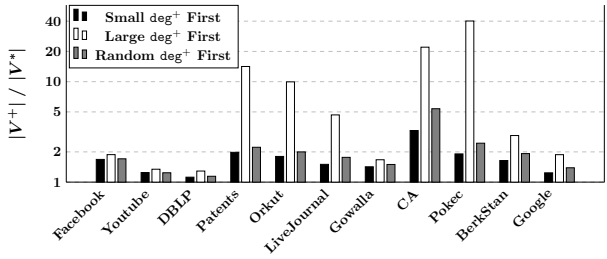,width=0.9\columnwidth}
\caption{Comparison of Heuristics to Generate the $k$-Order}
\label{fig:heuristic}
\end{figure}

\vspace{-0.1cm}
append $u$ to $O_{k-1}$; \quad $\rem(u) \leftarrow \TT{deg}(u)$; \\
\noindent \rule{\columnwidth}{1pt} \\
\noindent
In addition, we use a ``small $\rem$ first'' heuristic during the
generation process of $O_k$s, i.e., we always choose a vertex that has
the minimum $\rem$ to append to $O_{k-1}$. If there is a tie, we break
the tie arbitrarily. The justification is as follows: In Algorithm
\ref{alg:basic-insert}, to avoid a large $|V^+|$, the only way is to
reduce the probability of entering the first conditional branch (line
\ref{line:basic-insert-case1}). By putting vertices with smaller
$\rem$ before other vertices, when the algorithm visits the vertices
in order, it will first encounter those vertices with smaller
$\rem$. These vertices intuitively would be less likely to enter the
first branch, thus do not introduce new candidates to $V^+$, which in
return reduces the number of vertices to visit. We emphasize that the
algorithm with ``small $\rem$ first'' heuristic can still run in
linear time by adjusting the implementation by
\cite{batagelj:core-decomposition}.  To verify the effectiveness of our
heuristic, we compare it with another
two heuristics, namely,
``large $\rem$ first'' and ``random $\rem$ first''. By ``large $\rem$ first'',
we append to $O_{k-1}$ a vertex with largest $\rem$, while by
``random $\rem$ first'', we append to $O_{k-1}$ a vertex randomly as long as
its $\rem$ is smaller than $k$.
We show {\small $\frac{\NM{sum of }|V^+|\NM{ over all insertions}}
{\NM{sum of }|V^*|\NM{ over all insertions}}$} for three heuristics in
\figurename~\ref{fig:heuristic} after inserting 100,000 edges.  From the
figure, we can see that the ``small $\rem$ first'' heuristic 
consistently performs better in all
$11$ graphs tested.

\stitle{Implementation:}
In order to traverse $O_k$ efficiently, $O_k$ is implemented as a
doubly linked list.
Recall that we associate $O_k$ with a data
structure $\mathcal{A}_k$ to test $u \order v$ when both are in $O_k$.
In addition, we use a data structure $\mathcal{B}$ such that
jumping to $v_j$ (line
\ref{line:basic-insert-j} of Algorithm \ref{alg:basic-insert}) takes
$\mathcal{O}(1)$ time.
%
%We discuss three options
In order to make both \TT{\OI} (Algorithm \ref{alg:basic-insert}) and
\TT{\OR} (Algorithm \ref{alg:basic-removal}) efficient, there are
two main issues:
%
%
% \begin{itemize}[noitemsep,topsep=5pt]
% \item
% (a) how to support efficient vertices removals and insertions in $O_K$
% (e.g.,
% lines~\ref{line:basic-insert-remove}-\ref{line:basic-insert-insert} in
% Algorithm \ref{alg:basic-insert}), 
% \item
(A) how to implement $\mathcal{A}_k$, and 
% \item
(B) how to implement $\mathcal{B}$.
% \end{itemize}
%
% For (a), we implement every $O_K$ and $V_C$ using a doubly linked
% list. We focus on (b) and (c) below.

\stitle{(A) How to implement $\mathcal{A}_k$:}
% If $\TT{core}(u) \neq \TT{core}(v)$, whether $u \order v$ can be
% answered by testing whether $\TT{core}(u) < \TT{core}(v)$.
Here, the data structure is needed to test $u \order v$ when both are
in the same $O_k$; that is, $\TT{core}(u) = \TT{core}(v)$.  On selecting
an appropriate data structure for such a purpose, we take the
following observation into consideration.

\begin{observation} \label{obs:diff}
In algorithm \TT{\OI}, it is possible that $u \order v$
in the $k$-order $\order$ of $G$ is changed to $v \order' u$
in the new order $\order'$
after $G$ is updated by inserting/removing an edge.
\end{observation}

We explain how it happens. Consider processing vertices in $O_k$ from
left to right one-by-one as an
example. Suppose $u \order v$.
Initially, suppose $u$ is in {\bf Case-1}. As a result, by the
\TT{\OI} algorithm, $u$ is appended to $V_C$ instead of $O'_K$ as a
candidate to be in $V^*$. While $u$ is in $V_C$, the \TT{\OI}
algorithm processes $v$, finds that $v$ is in {\bf Case-2}, and thus
appends $v$ to $O'_K$. Some time later, it is possible that $u$ is
found not to be a candidate any more when the algorithm processes a
vertex $w$, which is in {\bf Case-2b}. Note that $w$ may be $v$.  Then
$u$ is removed from $V_C$ and also appended to $O'_K$.  This causes $v
\order' u$ in the new $k$-order $\order'$.  We emphasize that there
are at most $|V^+|$ such $u$s on a single edge insertion because $V^+$
includes all vertices that are in {\bf Case-1} in the \TT{\OI}
algorithm. We discuss how to support $\mathcal{A}_k$ below.

In current implementation, we encode $\mathcal{A}_k$ using an
\textit{order statistics tree}, where each tree node holds exactly one vertex.
The following invariant is maintained: for any two vertices $u$ and
$v$ in $O_k$, if $u \order v$, then
(1) $u$ is in the left subtree of $v$, or
(2) $v$ is in the right subtree of $u$, or
(3) there exists $w \in O_k$ such that $u$ is in the left subtree of $w$
and $v$ is in the right subtree of $w$.
By associating each node with the size of the subtree rooted at the
node, we are able to find the rank of a vertex in $O_k$ in
$\mathcal{O}(\log |O_k|)$ time \cite{cormen:introduction}. Such a
method faces an issue.
%
%, which highlights the difference between the trees we use and
%ordinary order statistics trees.
%
Given an order statistics tree, in order to find the rank of a vertex,
we need to first locate the node containing the vertex, starting from
the root. But we cannot decide which child pointer to follow to find
the node, since we do not have the rank information yet.  In other
words, given a vertex, in order to get the rank of the vertex using
the tree, we should first know the rank of the vertex.
%
% We can not decide it, after all the vertices are not
% ordered by their IDs.
% \textit{It is a chicken and egg problem.}
%
% \end{enumerate}
%
We propose a mechanism to address this issue by additionally creating a
one-to-one mapping between vertices in $O_k$ and nodes in the order
statistics tree. In this way, locating the target node in the order
statistics tree can be done easily.
It is worth noting that in our algorithm, vertices are inserted either
to the beginning of $O_{K+1}$ for insertion or the end of $O_{K-1}$ for
removal. Accordingly, we only need to follow either the left child
pointers or the right child pointers to insert the vertex to the order
statistics tree. Therefore, the issue mentioned above is avoided.
%
% followed by a sequence of local rotations.
%
The post-processing to make the tree balanced is similar to that of an
ordinary tree.
Both insertion and removal of a vertex takes $\mathcal{O}(\log |O_k|)$
time.

We show below that $\mathcal{A}_k$s
deal with Observation \ref{obs:diff} well.  To make it consistent
with the new $k$-order $O'_k$, we adjust the position of $u$ (refer to
Observation \ref{obs:diff}) in the order statistics tree as follows.
First, we remove $u$ from the tree. Suppose $u'$ is the previous
vertex of $u$ in $O'_k$. We insert $u$ to the right subtree of $u'$ by
following only left child pointers such that $u$ is the successor of
$u'$ in the tree. Local rotations are then taken to make the tree
valid. The total cost for adjusting positions thus is
$\mathcal{O}(|V^+|\log |O_k|)$ because there are at most $|V^+|$ such
$u$s and each takes $\mathcal{O}(\log |O_k|)$ time.
This additional cost does not affect the complexity of Theorem
\ref{thm:insertion-complexity}.

\begin{theorem} \label{thm:space}
The total space cost of $\mathcal{A}_k$s is $\mathcal{O}(n)$ and
for each $k$,  $\mathcal{A}_k$ can be created in $\mathcal{O}(|O_k|\log |O_k|)$
time by inserting vertices in $O_k$ one by one.
\end{theorem}

% Formally, $\mathcal{T}_k$s support (1)
% \TT{TreeRank}($u$): return the rank of $u$; (2) \TT{TreeRemove}($u$): remove $u$
% from the tree containing it; (3) \TT{TreeInsertFront}($u$, $\mathcal{T}$):
% insert $u$ to $\mathcal{T}$ as if it is the minimum element;
% (4) \TT{TreeInsertBack}($u$, $\mathcal{T}$): insert $u$ to $\mathcal{T}$
% as if it is the maximum element; (5) \TT{TreeInsertAfter}($u$, $v$): insert $u$ to
% the tree containing $v$ such that $v$ is the predecessor of $u$ after insertion.
% The last method is used for adjusting the position of a vertex in the tree.
% All these operations are easy to implement and we will not describe them in
% details here. They all run in $\mathcal{O}(\log|O_k|)$ time and
%  $\mathcal{T}_k$'s totally occupy $\mathcal{O}(n)$ space.

\vspace{-0.1cm}
 
\stitle{(B) How to implement $\mathcal{B}$:} $\mathcal{B}$ is
implemented as a min-heap. $\mathcal{B}$ maintains a set of
($\TT{rank}(w)$, $w$) pairs with $\ext(w) \neq 0$ or $\rem(w)$ $ > K$
({\bf Case-1} and {\bf Case-2b}), and uses the rank as the key.  When
we need to find the $v_j$ (line \ref{line:basic-insert-j} of Algorithm
\ref{alg:basic-insert}), the top pair of $\mathcal{B}$ is returned in
$\mathcal{O}(1)$ time. It requires $\mathcal{O}(\log |\mathcal{B}|) =
\mathcal{O}(\log |O_K|)$ time to insert a pair to $\mathcal{B}$ since
only vertices in $O_K$ will be inserted to $\mathcal{B}$.  It requires
$\mathcal{O}(\log |\mathcal{B}|)$ time to remove arbitrary pair from
$\mathcal{B}$.

\vspace{-0.3cm}

\section{Performance Studies} \label{sec:exp}

We have conducted experimental studies using 11 real large graphs, and we
report the performance of our algorithms by comparing with the
traversal algorithms. All algorithms are implemented in C++ and
compiled by g++ compiler at -O2 optimization level. For the new
order-based algorithms, we implement the order statistics tree on
top of \textit{treaps}. For the traversal algorithms,
we apply the enhancement proposed in \cite{sariyuce:streaming_journal}, which
is the journal version of \cite{sariyuce:streaming}. The enhancement consists
of exploiting neighborhood of higher hops to improve the pruning power of
$\TT{pcd}$, which considers only $2$-hop neighborhood of a vertex, at a higher
maintenance cost. Note that the traversal removal algorithm requires only
$\TT{mcd}$ and thus its performance degrades for higher hops.
We set the hop count $h$ in the range
$\{2, 3, 4, 5, 6\}$ as in \cite{sariyuce:streaming_journal} and denote each
version as $\TT{\TRA}\textnormal{-}h$.
% 
% Note that our choice
% does not mean that treaps are most suitable for our algorithms.  There
% may be more efficient ones but selecting the best one is out of scope
% of this paper.
%
All experiments are conducted on a Linux machine with Intel i7-4790
CPU and 32 GB main memory.

\begin{figure}[t]
\begin{subfigure}{0.235\textwidth}
\centering
\epsfig{file=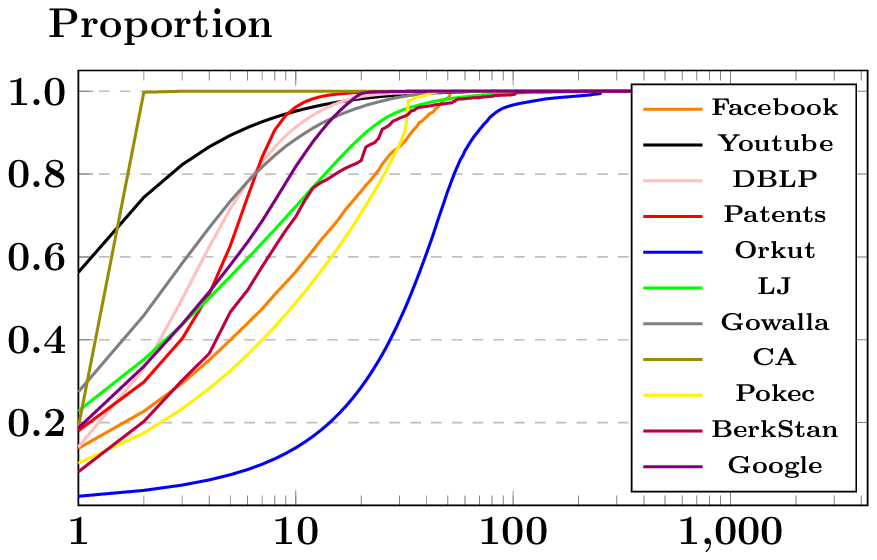,width=0.95\columnwidth}
\caption{The cumulative distribution of core numbers}
\label{fig:data-core-dist}
\end{subfigure}
\ 
\begin{subfigure}{0.235\textwidth}
\centering
\epsfig{file=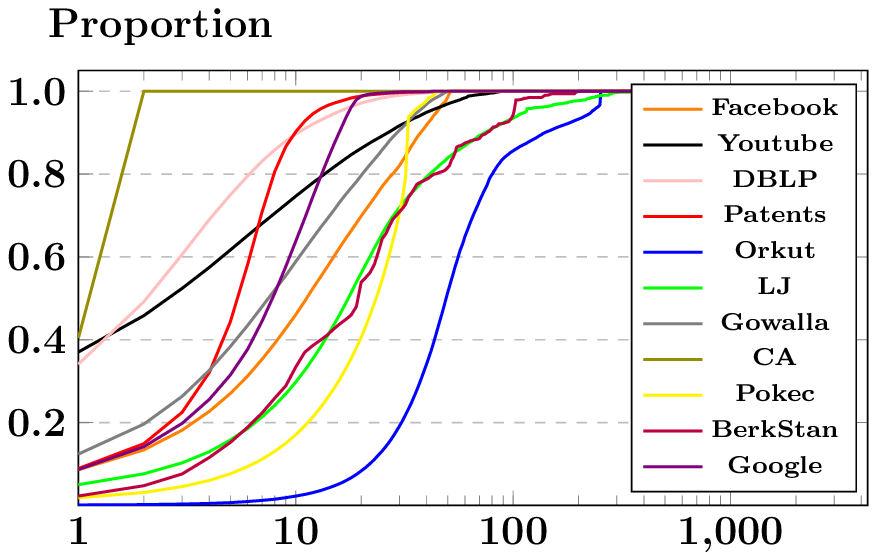,width=0.95\columnwidth}
\caption{Cumulative distribution of $K$ of the 100,000 edges sampled}
\label{fig:query-core-dist}
\end{subfigure}
\caption{Statistics of the Graphs and Edges Tested}
\end{figure}

\stitle{Datasets:} We use $11$ datasets publicly accessible, which are
shown in Table \ref{tbl:dataset}. These real datasets cover a wide
range of graphs used in different applications with different
properties.  \TT{Facebook}, \TT{Youtube}, and \TT{DBLP} are three
temporal graphs, whose edges are labeled with time stamps to show when they are
inserted, and can be downloaded from Konect ({\small
  \url{http://konect.uni-koblenz.de}}). The remaining eight graphs,
which can be downloaded from SNAP ({\small
  \url{https://snap.stanford.edu}}), include Social Networks
(\TT{LiveJournal}, \TT{Pokec}, and \TT{Orkut}), Citation Network (\TT{Patents}),
Web Graphs (\TT{BerkStan} and \TT{Google}), Road Network (\TT{CA}), and
Location-Based Social Network (\TT{Gowalla}). Directed graphs are converted to
undirected ones in our testing. 
%
% such that there is an edge between two vertices in the new graph if
% and only if there is at least one directed edge between the vertices
% in the original graph.
%
The statistics are shown in Table \ref{tbl:dataset}.  The cumulative
distribution of core numbers for each graph is shown in
\figurename~\ref{fig:data-core-dist}.

%\comment{
\begin{table}
\centering
{\footnotesize
\begin{tabular}{|l||r|r|r|r|}
\hline
Dataset     & $n = |V|$     & $m = |E|$    & avg. \TT{deg} & max $k$ \\
\hline\hline
\TT{Facebook}    & 63,731   & 817,035    & 25.64       & 52       \\ \hline
\TT{Youtube}     & 3,223,589 & 9,375,374   & 5.82        & 88       \\ \hline
\TT{DBLP}        & 1,314,050 & 5,362,414   & 8.16        & 118      \\ \hline
\TT{Patents}     & 3,774,768 & 16,518,947  & 8.75        & 64       \\ \hline
\TT{Orkut}       & 3,072,441 & 117,185,083 & 76.28       & 253      \\ \hline
\TT{LiveJournal} & 4,846,609 & 42,851,237  & 17.68       & 372      \\ \hline
\TT{Gowalla}     & 196,591  & 950,327    & 9.67        & 51       \\ \hline
\TT{CA}          & 1,965,206 & 2,766,607   & 2.82        & 3        \\ \hline
\TT{Pokec}       & 1,632,803 & 22,301,964  & 27.32       & 47       \\ \hline
\TT{BerkStan}    & 685,230  & 6,649,470   & 19.41       & 201      \\ \hline
\TT{Google}      & 875,713  & 4,322,051   & 9.87        & 44       \\ \hline
\end{tabular}%
}
\caption{Datasets} \label{tbl:dataset}
\end{table}
%}

\comment{
\begin{table}
\centering
\scalebox{0.7}{
\begin{tabular}{|l||r|r|r|r|r|r|}
\hline
Dataset     & $n = |V|$     & $m = |E|$    & avg. \TT{deg} & max $k$ & avg.
\TT{oc} & avg. \TT{pc} \\
\hline\hline 
\TT{Facebook}    & 63,731     & 817,035     & 25.64       & 52    & 17.40   &
77.07 \\ \hline
\TT{Youtube}     & 3,223,589  & 9,375,374   & 5.82        & 88    & 1.69    &
2.72 \\ \hline
\TT{DBLP}        & 1,314,050  & 5,362,414   & 8.16        & 118   & 6.70    &
364.24 \\ \hline
\TT{Patents}     & 3,774,768  & 16,518,947  & 8.75        & 64    & 517.50  &
71107.91 \\ \hline
\TT{Orkut}       & 3,072,441  & 117,185,083 & 76.28       & 253   & 337.93  &
6766.42\\ \hline
\TT{LiveJournal} & 4,846,609  & 42,851,237  & 17.68       & 372   & 22.82   &
1974.41\\ \hline
\TT{Gowalla}     & 196,591    & 950,327     & 9.67        & 51    & 2.75    &
17.69 \\ \hline
\TT{CA}          & 1,965,206  & 2,766,607   & 2.82        & 3     & 18.73   &
858013.22\\ \hline
\TT{Pokec}       & 1,632,803  & 22,301,964  & 27.32       & 47    & 3409.04 &
9660.40\\ \hline
\TT{BerkStan}    & 685,230    & 6,649,470   & 19.41       & 201   & 22.68   &
269.34 \\ \hline
\TT{Google}      & 875,713    & 4,322,051   & 9.87        & 44    & 4.51    &
94.11 \\ \hline \end{tabular}%
}
\caption{Datasets} \label{tbl:dataset}
\end{table}
}

\subsection{Order-Based vs Traversal}

We compare the performance of our order-based algorithms with the
traversal algorithms, which are the state-of-the-art.
Here, we select 100,000 edges out of each of the 11 graphs as follows.
For the three graphs with time stamps on edges, \TT{Facebook}, \TT{Youtube} and
\TT{DBLP}, because each edge is associated with a time stamp, we select the
latest 100,000 edges, i.e., the edges with maximum time stamps.  For
each of the remaining eight graphs, we randomly sample 100,000 edges.
For each graph, we show the cumulative distribution of $K$ of the 100,000
edges sampled in \figurename~\ref{fig:query-core-dist}, where
$K = \min\{\TT{core}(u), \TT{core}(v)\}$ given an edge $(u, v)$.
\figurename~\ref{fig:query-core-dist} shows that the edges to test cover
$k$-cores in wide range, implying they are suitable for testing.
For each graph, we measure the accumulated time for inserting these 100,000
edges one by one, and then measure the accumulated time for removing these
edges from the graph. The results are shown in \tablename~\ref{tbl:perf}.

\vspace{-0.1cm}
\stitle{Edge Insertion:} We compare \TT{\OI} (Algorithm
\ref{alg:basic-insert}) with the traversal insertion algorithm
\cite{sariyuce:streaming} (\cite{sariyuce:streaming_journal}). The results are
shown in \tablename~\ref{tbl:perf}.  \TT{\OI} significantly outperforms the
traversal insertion algorithm in all the datasets tested. For \TT{Pokec},
the speedup even achieves up to 2,083 times. For \TT{Patents} and \TT{CA},
we further test with higher $h$ for the traversal insertion algorithm
to get the optimal efficiency and the results are 810.00s ($h = 6$) for
\TT{Patents} and 1.08s ($h = 7$) for \TT{CA}, which are still slower than
\TT{\OI}.
As shown in
\figurename~\ref{fig:perf-expansion}, \TT{\OI} visits much less 
vertices than $\TT{\TRA}\textnormal{-}2$. This confirms
the efficiency of \TT{\OI} algorithm.
In addition, we emphasize that \TT{\OI} has much smaller performance
variation between edges inserted, as shown in \figurename~\ref{fig:dist}.

\begin{table*}
\centering
\begin{tabular}{|l||r|r|r|r|r|r||r|r|r|r|r|r|}
\hline
& \multicolumn{6}{c||}{\textbf{Insert}} & \multicolumn{6}{c|}{\textbf{Remove}}
\\
\hline
\textbf{Dataset} & \TT{\OI} & \TT{\TRA}-2 & \TT{\TRA}-3 & \TT{\TRA}-4 &
\TT{\TRA}-5 & \TT{\TRA}-6 & \TT{\OR} & \TT{\TRA}-2 & \TT{\TRA}-3 & \TT{\TRA}-4
& \TT{\TRA}-5 & \TT{\TRA}-6 \\
\hline\hline

\TT{Facebook} & \fastest{0.16} & \runnerup{3.52} & 4.07 & 5.91 & 10.52 & 16.95
& \fastest{0.10} & \runnerup{0.50} & 1.63 & 4.14 & 9.70 & 17.77 \\ \hline
\TT{Youtube} & \fastest{0.26} & \runnerup{2.51} & 2.88 & 4.01 & 6.13 & 9.71
& \fastest{0.28} & \runnerup{0.61} & 1.42 & 3.19 & 6.28 & 11.32\\ \hline
\TT{DBLP} & \fastest{0.16} & 1.80 & \runnerup{1.20} & 2.31 & 6.32 & 17.65
& \fastest{0.11} & \runnerup{0.21} & 0.61 & 1.88 & 5.49 & 15.78 \\ \hline
\TT{Patents} & \fastest{0.88} & 2,944.14 & 1,805.98 & 1,173.20 & 845.93 &
\runnerup{810.00}
& \fastest{0.38} & \runnerup{0.92} & 4.22 & 18.57 & 75.06 & 276.37
\\ \hline 
\TT{Orkut} & \fastest{1.14} & 954.36 & 793.82 & \runnerup{780.69} & 996.43 &
1,576.63
& \fastest{0.71} & \runnerup{7.75} & 36.80 & 136.78 & 428.85 & 1,089.38
\\ \hline
\TT{LiveJournal} & \fastest{0.53} & 149.56 & 90.93 & \runnerup{76.57} & 125.29 &
285.50
& \fastest{0.33} & \runnerup{1.66} & 6.59 & 24.56 & 86.10 & 233.92
\\ \hline
\TT{Gowalla} & \fastest{0.18} & \runnerup{1.04} & 1.37 & 2.21 & 3.78 & 6.38
& \fastest{0.14} & \runnerup{0.35} & 0.84 & 1.82 & 3.45 & 6.22
\\ \hline
\TT{CA} & \fastest{0.52} & 15.14 & 4.20 & 2.08 & 1.37 & \runnerup{1.11}
& 0.16 & \fastest{0.08} & \runnerup{0.13} & 0.19 & 0.26 & 0.33
\\ \hline
\TT{Pokec} & \fastest{0.77} & 1,726.04 & \runnerup{1,603.80} & 1,650.37 &
1,876.48 & 2,338.78
& \fastest{0.32} & \runnerup{4.86} & 53.13 & 259.93 & 756.40 & 1,652.88
\\\hline
\TT{BerkStan} & \fastest{0.37} & \runnerup{6.37} & 7.29 & 9.37 & 13.14 & 16.19
& \fastest{0.52} & \runnerup{2.55} & 5.04 & 8.33 & 12.45 & 17.34
\\ \hline
\TT{Google} & \fastest{0.37} & \runnerup{1.01} & 1.25 & 2.44 & 4.81 & 9.27
& \fastest{0.25} & \runnerup{0.46} & 0.96 & 2.08 & 4.32 & 8.75
\\ \hline
\end{tabular}
\caption{Performance Comparison (in seconds): Order-Based vs Traversal
(The winner is in {\color{red}{\textbf{bold}}}; the runner-up is in
{\color{blue}{\textit{italics}}})}
\label{tbl:perf}
\vspace{-3em}
\end{table*}

\vspace{-0.1cm}
\stitle{Edge Removal:} We compare \TT{\OR} (Algorithm
\ref{alg:basic-removal}) with the traversal removal algorithm
\cite{sariyuce:streaming} (\cite{sariyuce:streaming_journal}).  The results are
shown in \tablename~\ref{tbl:perf}. We observe that $\TT{\TRA}\textnormal{-}2$ is
more efficient than those with higher $h$, and \TT{\OR} outperforms
$\TT{\TRA}\textnormal{-}2$ in all datasets except \TT{CA}. Here,
the average degree of \TT{CA} is only $2.82~{}(\approx 3)$, as shown in
\tablename~\ref{tbl:dataset}. Such a small average degree indicates that the
cost to maintain $\TT{pcd}$ is low, because the number of vertices whose $\TT{pcd}$ need to be
updated after an edge update is small.  On the other hand, \TT{\OR}
outperforms $\TT{\TRA}\textnormal{-}2$
in all other graphs tested, especially \TT{Pokec} and \TT{Orkut},
because \TT{Pokec} and \TT{Orkut} have the highest
average degrees among all the datasets.

\vspace{-0.1cm}
\stitle{Index Space and Index Creation:} Our
order-based algorithm requires $\mathcal{O}(n)$ space,
as shown in Theorem~\ref{thm:space}.
In our current implementation, the order-based algorithm needs about
five times of the space used by $\TT{\TRA}\textnormal{-}2$
and needs only about double time to create the index (including computing the
initial core numbers).
When the hop count $h$ becomes larger, the space and time required by the
traversal algorithms also become higher. See \tablename~\ref{tbl:index}.
\comment{
For all datasets tested, we also compare the time to create index using the
ratio {\small $\frac{\NM{index creation time of the order-based
    algorithm}}{\NM{index creation time of the traversal algorithm}}$}.
The results are shown in \figurename~\ref{fig:perf-creation}. As can be
seen, the order-based algorithm needs only about double time to create the
index.
}
We emphasize that index creation is a one-time cost. Moreover, in view
of the insertion/removal performance improvement brought, such a tradeoff
is appropriate.

\begin{figure}[t]
\begin{subfigure}{0.235\textwidth}
\epsfig{file=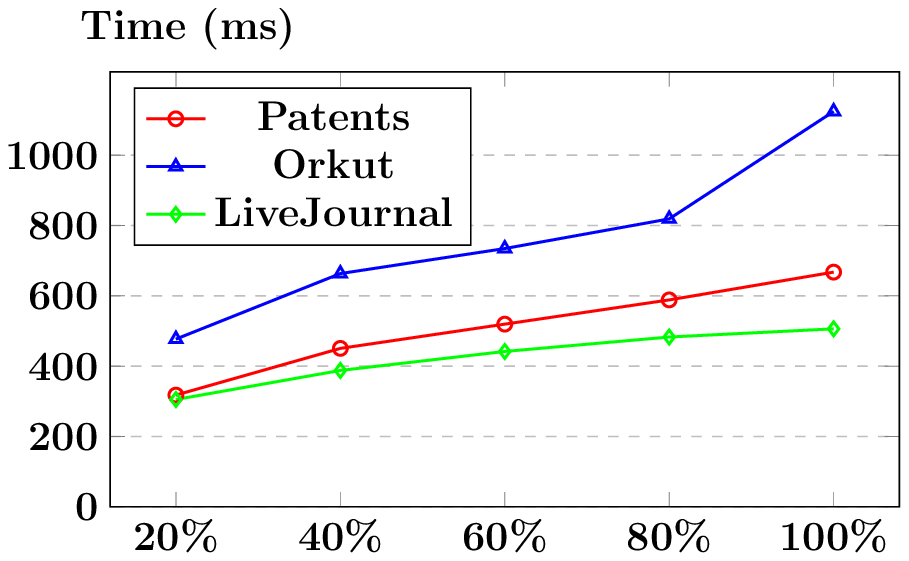,width=0.9\columnwidth}
\caption{Vary $|V|$: time} \label{fig:scalability-V-time}
\end{subfigure}
\begin{subfigure}{0.235\textwidth}
\epsfig{file=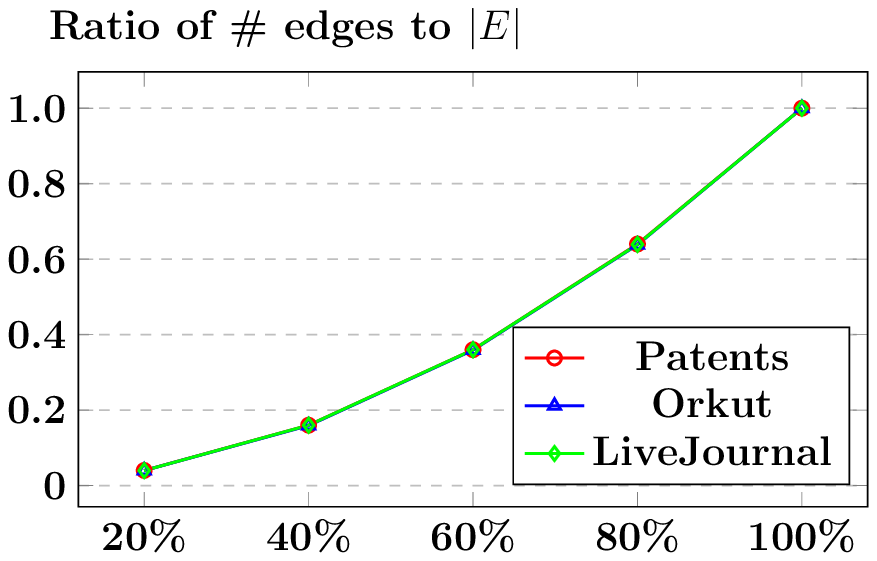,width=0.9\columnwidth}
\caption{Vary $|V|$: edge ratio} \label{fig:scalability-V-ratio}
\end{subfigure}
\begin{subfigure}{0.235\textwidth}
\epsfig{file=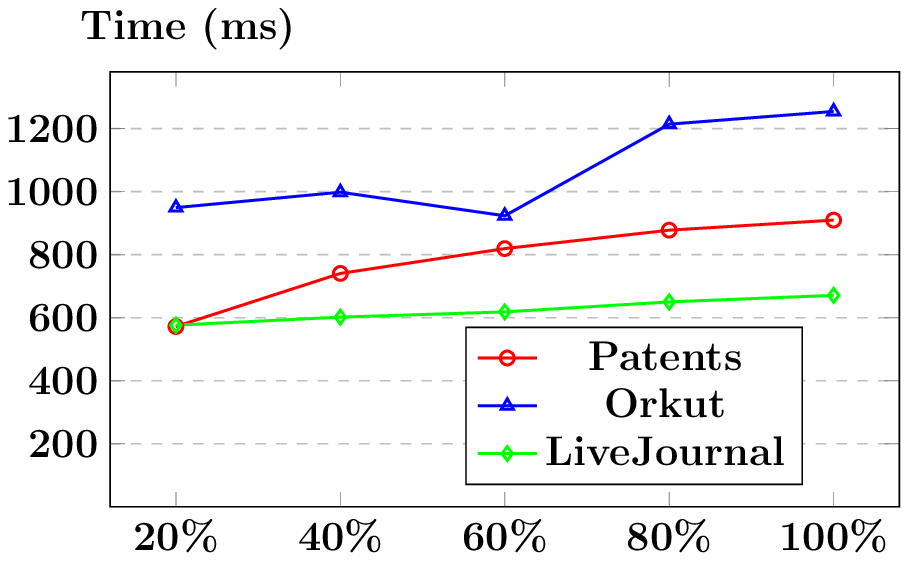,width=0.9\columnwidth}
\caption{Vary $|E|$: time} \label{fig:scalability-E-time}
\end{subfigure}
\begin{subfigure}{0.235\textwidth}
\epsfig{file=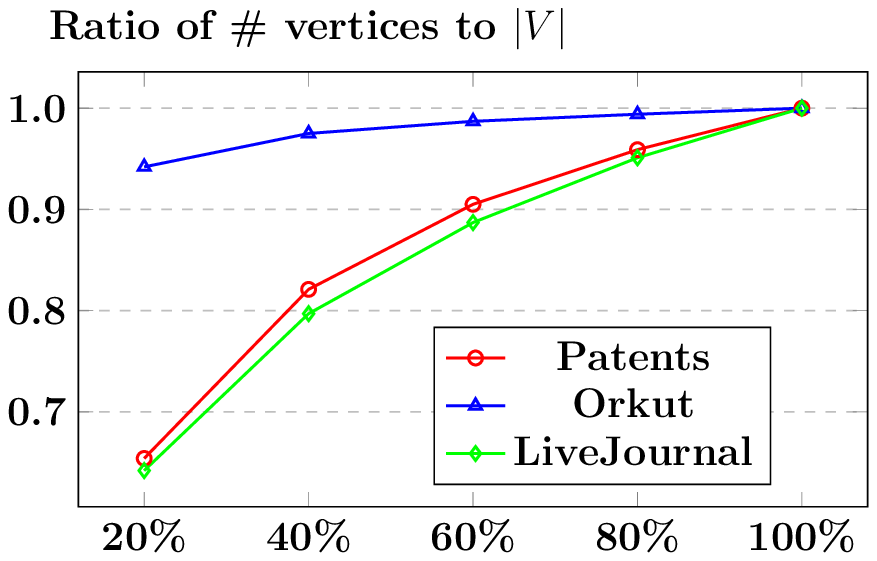,width=0.9\columnwidth}
\caption{Vary $|E|$: vertex ratio} \label{fig:scalability-E-ratio}
\end{subfigure}
\caption{The Scalability of the \TT{\OI} Algorithm} \label{fig:scalability}
\end{figure}

\subsection{Scalability Testing}

% We conduct testing to show \TT{\OI}'s performance does not degrade
% when the graph becomes larger.
%
We test the scalability of \TT{\OI} using the largest three datasets,
i.e., \TT{Patents}, \TT{Orkut}, and \TT{LiveJournal}.  We vary the number of
vertices $|V|$ and the number of edges $|E|$ respectively by randomly sampling
vertices and edges at rates from 20\% to 100\%. When sampling
vertices, we use the subgraph induced by the vertices.  When sampling
edges, we keep the incident vertices of the edges. From each subgraph sampled,
we further sample 100,000 edges for testing. We
show the results in \figurename~\ref{fig:scalability}.
\figurename~\ref{fig:scalability-V-time} shows the total time (ms) to insert
the sampled 100,000 edges one by one and update core numbers, and
\figurename~\ref{fig:scalability-V-ratio} shows the ratio of the number of
edges in each sampled subgraph to the number of edges in the original
graph, while varying $|V|$.  As can be seen, the time taken by
\TT{\OI} grows smoothly while the number of edges increase rapidly.
\figurename~\ref{fig:scalability-E-time} shows the total time (ms) to
insert the sampled 100,000 edges and update core numbers,
and \figurename~\ref{fig:scalability-E-ratio} shows the ratio of the number of
vertices in each sampled subgraph to the number of vertices in the
original graph, while varying $|E|$.  Similarly, \TT{\OI}
performs well while the number of vertices increase rapidly.
\comment{
Observe that when the sample rate is 60\%, there is a pit for \TT{Orkut} in
\figurename~\ref{fig:scalability-E-time}. This is because the total number
of vertices whose core numbers need to be updated in this case happens
to be slightly less than that in case 40\% and 80\%.  As a result, our
algorithm visits less vertices.  In summary, \TT{\OI} shows high
scalability and thus is suitable for inserting edges for graphs of
different size.
}
We do not show the scalability results for \TT{\OR} because it
only relies on $V^*$ and their neighbors, as shown in
Theorem~\ref{thm:removal-complexity}.  The scalability thus is not an issue
of \TT{\OR}.

\subsection{Stability Testing}

\begin{figure}[t]
\begin{subfigure}{0.235\textwidth}
\epsfig{file=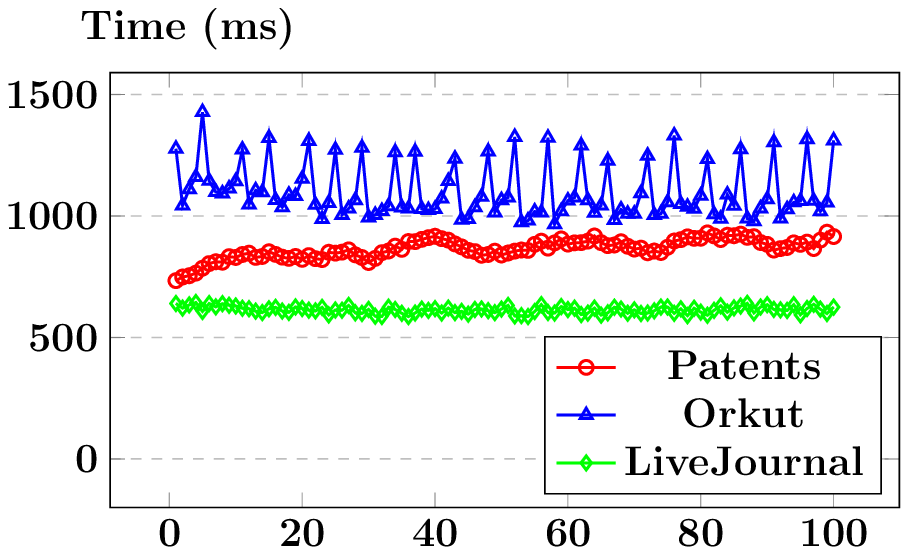,width=0.9\columnwidth}
\caption{$p = 0$} \label{fig:stab-0}
\end{subfigure}
\begin{subfigure}{0.235\textwidth}
\epsfig{file=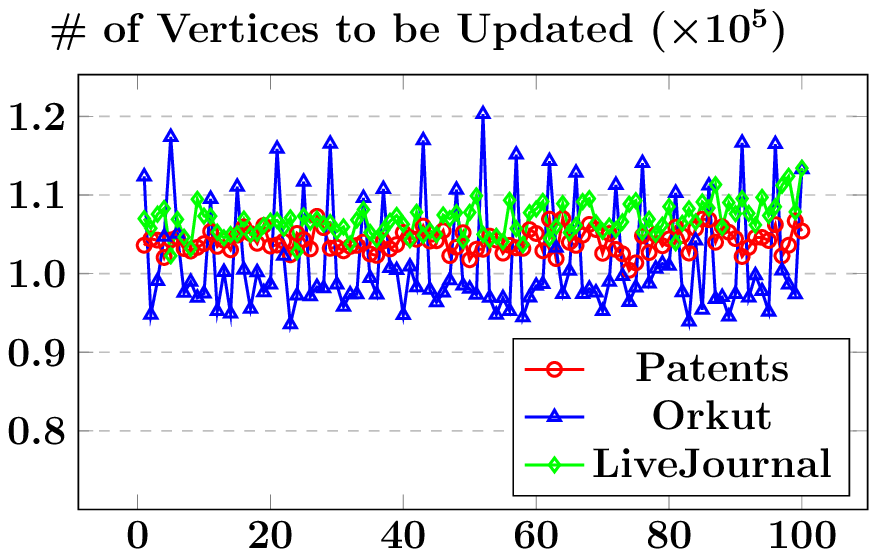,width=0.9\columnwidth}
\caption{\# of Vertices to be Updated} \label{fig:stab-cnt}
\end{subfigure}
\begin{subfigure}{0.235\textwidth}
\epsfig{file=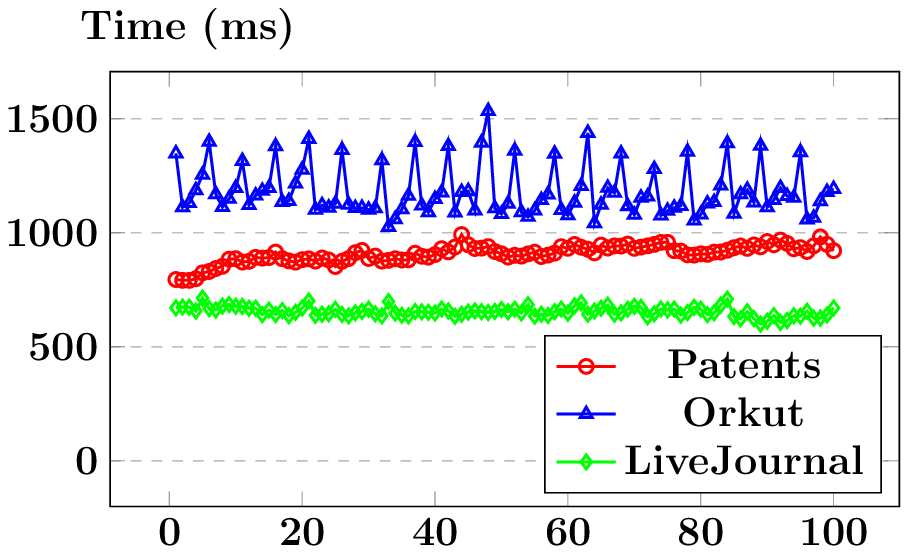,width=0.9\columnwidth}
\caption{$p = 0.1$} \label{fig:stab-10}
\end{subfigure}
\begin{subfigure}{0.235\textwidth}
\epsfig{file=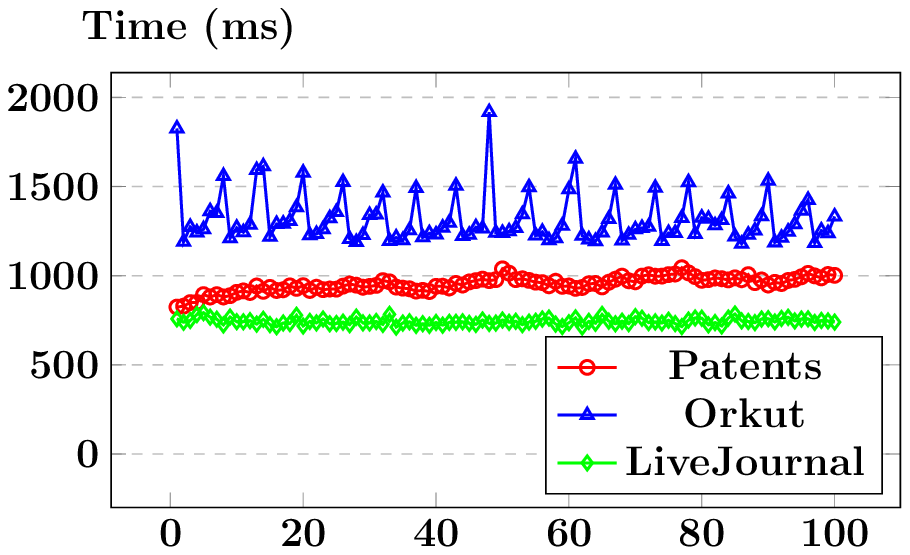,width=0.9\columnwidth}
\caption{$p = 0.2$} \label{fig:stab-20}
\end{subfigure}
\vspace{-0.2cm}
\caption{The Stability of the \TT{\OI} Algorithm}\label{fig:stability}
\end{figure}

\vspace{-0.1cm}
We update the $k$-order when an edge inserted. An issue is whether the
effectiveness of the $k$-order will be stable after a large number of
edge insertions.
We test the stability of the algorithm using the largest three
datasets, i.e., \TT{Patents}, \TT{Orkut}, and \TT{LiveJournal}.  The stability
testing is conducted as follows. First, we randomly sample 10,000,000 edges
from the graph and randomly partition them into 100 groups, where each
group has 100,000 edges. Second, we reinsert these edges to the graph
group by group. For each group, we measure the accumulated time used by \TT{\OI}
to insert the edges one by one. \figurename~\ref{fig:stab-0} shows the
results. The performance of the \TT{\OI} insertion algorithm is well
bounded, for all three datasets. For \TT{Orkut} graph, we observe that its
result fluctuates frequently. This is due to the fact that the number of
vertices whose core numbers need to be updated inside each group varies a
lot, as can be seen in \figurename~\ref{fig:stab-cnt}. To make the
experiment more practical, we conducted two additional experiments as
follows. For each group, after an edge is inserted, with probability
$p$ we randomly remove an edge from the graph. In the experiments, we
set $p$ as $0.1$ and $0.2$ respectively and the total time measured also
includes the time of removing edges. The results are similar to
\figurename~\ref{fig:stab-0}, as shown in \figurename~\ref{fig:stab-10} and
\ref{fig:stab-20}.

\vspace{-0.2cm}
\section{Conclusions}
\label{sec:con}

In this paper, we study core maintenance to reduce the computational
cost to compute $k$-cores for a graph when edges are inserted/removed
dynamically. To further improve the efficiency of the state-of-the-art
traversal algorithm, we propose a new order-based algorithm. The key
is to explicitly maintain the $k$-order for a graph. With the
$k$-order, we can identify the set of vertices that need to be updated
efficiently, and we can maintain the $k$-order for a graph with small
overhead when edges are inserted/removed. We confirm our
approach by conducting extensive performance studies using 11 real large graphs.
Our \TT{\OI} algorithm outperforms the traversal insertion algorithm
up to 3 orders of magnitude.

\comment{
\section*{Acknowledgment}
The work was supported by grant of Research Grants
Council of the Hong Kong SAR, China No. 14209314.
}

%\end{document}  % This is where a 'short' article might terminate

% ensure same length columns on last page (might need two sub-sequent
% latex runs)
%\balance

%ACKNOWLEDGMENTS are optional
% \section{Acknowledgments}
% \nocite{*}

\vspace{-0.2cm}
{\small
\bibliographystyle{abbrv}
%\nocite{*}
\bibliography{IEEEabrv,ref}
}

\appendix[Proof of Theorems and Lemmas]

\stitle{Proof Sketch of Lemma \ref{lmm:rem}}
The \textit{if} part: In Algorithm
\ref{alg:core-decomposition}, if vertices are removed in the order
$O_0O_1O_2\cdots$, it is guaranteed that the remaining degree of a
vertex $v$ is $\rem(v) < k + 1$, when $v$ is the current vertex being
processed.  Thus, $v$ can be removed. The \textit{only if} part: this
is trivial and omitted.

\stitle{Proof Sketch of Lemma \ref{lmm:instant-termination}}
We have $\rem(w) \leq K$ for $\forall w \in O_K$
if $\rem(u) \leq K$.  Let $w$ be the first vertex in $O_K$. The number of
$w$'s neighbors in the $(K + 1)$-core of
$G'$ is at most $\rem(w) \leq K$. Thus, $w$ can not
be in the $(K + 1)$-core of $G'$.
We can infer similarly for remaining vertices in $O_K$.

\stitle{Proof Sketch of Lemma \ref{lmm:before-root}}
After $(u, v)$ is inserted, $\rem(w)$ for the vertex $w$
appearing before $u$ will remain unchanged, i.e., $\leq K$.
The proof is similar to that of Lemma \ref{lmm:instant-termination}
and is thus omitted.

\stitle{Proof Sketch of Theorem \ref{thm:insertion-complexity}}
The preparing phase is in $\mathcal{O}(\log |O_K|)$ time because we need
to test whether $u \order v$, using $\mathcal{A}_K$.
The main part of the core phase is the while loop
(lines~\ref{line:basic-insert-for-beg}-\ref{line:basic-insert-for-end}).
% We count how many times it will enter the while loop.
First, the algorithm enters the second branch
(lines~\ref{line:basic-insert-case2a}-\ref{line:basic-insert-case2a-end})
in the while loop at most $|V^+| + 1$ times.  Because
lines~\ref{line:basic-insert-case2a}-\ref{line:basic-insert-case2a-end}
take $\mathcal{O}(1)$ time using $\mathcal{B}$, the total time complexity for
this branch in the whole loop is $\mathcal{O}(|V^+|)$.
Second,
lines~\ref{line:basic-insert-case1}-\ref{line:basic-insert-case1-end}
take $\mathcal{O}(\TT{deg}(v_i)\log |O_K|)$ time, because we need to
enumerate neighbors of $v_i$ and for each neighbor $w$, we test $v_i
\order w$.  In total, this branch takes
$\mathcal{O}(\log|O_K|\cdot\sum_{v \in V^+}\TT{deg}(v))$ time during
the whole loop.
Third, the third branch
(lines~\ref{line:basic-insert-case2b}-\ref{line:basic-insert-case2b-end})
takes $\mathcal{O}(\sum_{v \in V^+}\TT{deg}(v)\log |O_K|)$ time in
total in the whole loop.  Note that Algorithm~\ref{alg:keep} is called
to remove more vertices from $V_C$ each time when the algorithm enters
this branch. Each vertex $w'$ that is removed from $V_C$ takes
$\mathcal{O}(\TT{deg}(w')\log |O_K|)$ time to update $\rem$ and $\ext$
and to find more vertices to remove. There are at most $|V^+|$ such
$w'$s in the whole loop because $w'$ must be in {\bf Case-1} for it to
be inserted to $V_C$.
For the ending phase,
because $V^* \subseteq V^+$, updating $\TT{mcd}$ takes
$\mathcal{O}(\sum_{v \in V^*}\TT{deg}(v)) = \mathcal{O}(\sum_{v \in
  V^+}\TT{deg}(v))$.
In addition,
it takes $\mathcal{O}(|V^*| \log \max\{|O_K|, |O_{K+1}|\})$ time
to move vertices between $\mathcal{A}_K$ and $\mathcal{A}_{K+1}$. Thus, the
complexity of Algorithm \ref{alg:basic-insert}
%
% and \ref{alg:keep}
%
is $\mathcal{O}((\sum_{v \in V^+}\TT{deg}(v))\cdot\log \max\{|O_K|,\linebreak 
|O_{K+1}|\})$.

\stitle{Proof Sketch of Theorem \ref{thm:basic-remove-list}}
For vertices that are still in $O_K$, their $\rem$s are not increased
in the algorithm. Thus, their $\rem$ remains smaller than or equal to
$K$. Let the vertices in $V^*$ be $w_1, w_2, \cdots, w_{|V^*|}$ in the
order they are inserted to $V^*$. The traversal removal algorithm
guarantees that when we insert a vertex $w_i$ to $V^*$, $\TT{deg}(w_i,
G(V_{\geq K} \setminus\{w_1, w_2, \cdots, w_{i-1}\})) \leq K - 1$,
where $V_{\geq K}$ is the original
set of vertices whose core numbers are not less than $K$.
Therefore, we have $\rem(w_i) \leq
K - 1$ for each $w_i \in V^*$.

\stitle{Proof Sketch of Theorem \ref{thm:removal-complexity}}
First, identifying $V^*$ takes only $\mathcal{O}(\sum_{w \in V^*}
\TT{deg}(w))$ time. Second, maintaining $O_K$ and $O_{K-1}$ requires
$\mathcal{O}(\log |O_K|\cdot \sum_{w \in V^*}\TT{deg}_K(w) +
|V^*|\cdot\log |O_{K-1}|)$ time.
We provide the analysis
below for each $w \in V^*$.
(a)
Lines~\ref{line:basic-remove-first-if-beg}-\ref{line:basic-remove-first-if-end}
totally take $\mathcal{O}(\TT{deg}(w) + \TT{deg}_K(w) \log |O_K|)$
time. Note that $w' \order w$ is evaluated only if $\TT{core}(w') =
K$.
(b)
Lines~\ref{line:basic-remove-second-if-beg}-\ref{line:basic-remove-second-if-end}
totally take $\mathcal{O}(\TT{deg}(w))$ time assuming testing $w' \in
V^*$ requires $\mathcal{O}(1)$ time.
(c)
Lines~\ref{line:basic-remove-post-beg}-\ref{line:basic-remove-post-end}
require $\mathcal{O}(1) + \mathcal{O}(\log |O_K| + \log |O_{K-1}|)$
time, where $\mathcal{O}(1)$ is for removing $w$ from $V^*$ and moving
$w$ from $O_K$ to $O_{K-1}$, while $\mathcal{O}(\log |O_K| + \log
|O_{K-1}|)$ is for removing $w$ from $\mathcal{A}_K$ and inserting $w$
to $\mathcal{A}_{K-1}$.
In addition, line~\ref{line:basic-remove-mcd} requires $\mathcal{O}(\sum_{v
\in V^*}\TT{deg}(v))$ time.

\appendix[Index Creation Time Comparison]
\begin{table}[h]
\centering
{\tiny
\begin{tabular}{|l||r|r|r|r|r|r|}
\hline
\textbf{Dataset} & Order-Based & \TT{\TRA}-2 & \TT{\TRA}-3 & \TT{\TRA}-4 &
\TT{\TRA}-5 & \TT{\TRA}-6 \\ \hline
\TT{Facebook} & 0.03 & 0.03 & 0.03 & 0.04 & 0.05 & 0.06
\\ \hline
\TT{Youtube} & 1.45 & 0.96 & 1.14 & 1.33 & 1.50 & 1.69
\\ \hline
\TT{DBLP} & 0.74 & 0.54 & 0.65 & 0.77 & 0.85 & 0.95
\\ \hline
\TT{Patents} & 3.93 & 3.65 & 4.32 & 5.03 & 5.76 & 6.32
\\ \hline 
\TT{Orkut} & 7.25 & 7.39 & 9.16 & 10.90 & 12.91 & 14.49
\\ \hline
\TT{LiveJournal} & 4.26	& 3.75 & 4.63 & 5.53 & 6.23	& 7.14
\\ \hline
\TT{Gowalla} & 0.07	& 0.06 & 0.07 & 0.09 & 0.10	& 0.11
\\ \hline
\TT{CA} & 0.77 & 0.35 & 0.43 & 0.50 & 0.58 & 0.64
\\ \hline
\TT{Pokec} & 2.49 & 2.67 & 3.02 & 3.33 & 3.90 & 4.11
\\\hline
\TT{BerkStan} & 0.32 & 0.29 & 0.37 & 0.44 & 0.51 & 0.58
\\ \hline
\TT{Google} & 0.45 & 0.35 & 0.44 & 0.52 & 0.59 & 0.66
\\ \hline
\end{tabular}
}
\caption{Time to Create Index (in seconds)} \label{tbl:index}
\end{table}

\end{document}